\documentclass[symmetry,article,moreauthors,accept,pdftex]{Definitions/mdpi}

\usepackage{amsmath}
\usepackage{amssymb}
\usepackage{amsfonts}
\usepackage{graphicx}
\usepackage{color} 
\usepackage{hyperref}

\firstpage{1} 
\pubvolume{13}
\issuenum{3}
\articlenumber{522}
\pubyear{2021}
\copyrightyear{2021}
\externaleditor{{Academic Editor: Rami Ahmad El-Nabulsi} 
} 
\datereceived{28 February 2021} 
\dateaccepted{22 March 2021} 
\datepublished{23 March 2021} 
\hreflink{https://doi.org/10.3390/sym13030522}  

\pdfoutput=2
\usepackage{booktabs} 
\usepackage{multirow}
\usepackage{soul} 
\usepackage{microtype}

\usepackage{mathtools}
\usepackage{wasysym}
\usepackage{stmaryrd} 
\usepackage{MnSymbol} 

\DeclareMathOperator{\bfifH}{\textbf{\textit{H}}}

\setitemize{parsep=6pt,itemsep=0pt,leftmargin=*,labelsep=5.5mm,align=parleft}  
\setenumerate{parsep=6pt,itemsep=0pt,leftmargin=*,labelsep=5.5mm,align=parleft}
\setlist[description]{itemsep=0mm}   

\Title{Reparametrization Invariance and Some of the Key Properties of Physical Systems}

\TitleCitation{Reparametrization-\linebreak Invariance and Some of the Key Properties of Physical Systems}

\Author{Vesselin G. Gueorguiev
 $^{1,2,}$*\orcidB{} and Andre Maeder $^{3}$\orcidA{} }

\AuthorNames{Vesselin G. Gueorguiev and Andre Maeder}
\AuthorCitation{Gueorguiev, V.G.;\linebreak Maeder, A.}
\address{%
$^{1}$ \quad Institute for Advanced Physical Studies, Sofia 1784, Bulgaria\\
$^{2}$ \quad Ronin Institute for Independent Scholarship, 127 Haddon Pl., Montclair, NJ 07043, USA\\
$^{3}$ \quad Geneva Observatory, 
University of Geneva, Chemin des Maillettes 51, CH-1290 Sauverny, Switzerland; andre.maeder at unige.ch 
}

\corres{{Correspondence}: Vesselin at MailAPS dot org}

\date{\today}

\abstract{
In this paper, we argue in favor of first-order homogeneous Lagrangians in the velocities.
The relevant form of such Lagrangians is discussed and justified physically and geometrically.
Such Lagrangian systems possess Reparametrization Invariance (RI) and explain the 
observed common Arrow of Time as related to the non-negative mass for physical particles. 
The extended Hamiltonian formulation, which is generally covariant 
and applicable to reparametrization-invariant systems, is emphasized. 
The connection between the explicit form of the extended Hamiltonian $\boldsymbol{H}$
and the meaning of the process parameter $\lambda$ is illustrated.
The corresponding extended Hamiltonian $\boldsymbol{H}$ defines the classical phase space-time 
of the system via the Hamiltonian constraint $\boldsymbol{H}=0$ 
and guarantees that the Classical Hamiltonian $H$ corresponds to 
$p_{0}$---the energy of the particle when the coordinate time parametrization is chosen. 
The Schr\"odinger's equation and the principle of superposition of quantum 
states emerge naturally.  A connection is demonstrated between the positivity
of the energy $E=cp_{0}>0$ and the normalizability of the wave function
by using the extended Hamiltonian that is relevant for the proper-time
parametrization.}



\begin{document}


\section{Introduction}

It has taken thousands of years for natural philosophers and thinkers to arrive 
at the law of inertia, to accept it, and to turn it into a useful scientific paradigm by
overthrowing the ``obvious'' Aristotelian physics (due to Aristotle 384--322 BCE). 
The process has been slow and painful with occasional advances in its formulation and 
better understanding of the law of inertia by various predecessors of Newton, 
such as Avicenna (Ibn Sīnā 980--1037),  Galileo (1564--1642), 
who formulated the law of inertia for horizontal movement on the Earth, 
and later generalized by Descartes in his ``Discourse of the Method'' (1637). 
Until it finally finds its place as Newton's first principle: 
an object maintains its state, of rest or constant velocity propagation through space,
unless a force acts on it, along with the fertile company of the other two laws whose
mathematical formulation has been a breakthrough in mechanics.

At first sight, such principle seems to
be untrue due to our everyday experience which shows that for an object
to maintain its constant velocity an external influence is needed.
The accumulation of knowledge and technological progress have made
it possible for Newton to find the framework and formulate the three
main principles that are now the cornerstone of Newtonian Mechanics.

In Newtonian Mechanics, time 
is a parameter that all observers that are connected via Galilean transformations
will find to be the same---as long as they use the same identical
clocks to keep a record of their time. The Galilean transformations
are reflecting the symmetry under which the lows of the Newtonian
Mechanics are form-invariant \cite{Anderson2017,Goldstain1980}. The
spatial coordinates of the processes studied may have different values
for different inertial observers, but these observers can compare
their observations and would find an agreement upon utilization of
the Galilean transformations. In this sense, the time coordinate is
disconnected/disjoint from the configuration space $M$, which is
used to label the states of the system/process, however, it is essential
for the definition of the velocity vectors in the cotangent space
$TM$.

In Special Relativity (SR), time 
becomes related to the observer and the Lorentz transformations intertwine
space and time together in a Minkowski space-time
\cite{Anderson2017, 2005LRR.....8....5M, Pauli1958}.
This way the time duration of a process could be measured by different
observers to be different even if they use identical laboratory clocks.
However, all observers can identify a time duration related to an
observer that is at rest with respect to the process's coordinate frame
(co-moving frame). This is the proper-time 
duration of a process. Then all observers that are connected by Lorentz
transformations will arrive at the same value for the proper-time
duration of a process. Special Relativity unifies the time coordinate
with the spatial coordinates of an observer to a spacetime
the configuration space of the coordinates of events.
This way, from the point of view of an observer, the space-time is
divided into three important subsets: the time-like paths, space-like
curves, and light-like paths or equivalently into a past and future
cones inside the light-cone defined by the light-like paths connected
to the observer, and the space-like exterior of the rest of space-time.
Lorentz transformations preserve the local light-cone at any point
in the space-time and thus the causal structure of the time-like paths
describing a physical process.

General Relativity (GR) goes even further by allowing comparison between
observers related by any coordinate transformations, as long as there
is an equivalent local observer who's space-time is of Minkowski type.
This means that time records associated with identical clocks that
undergo arbitrary physically acceptable motion/process can be compared
successfully---that is, the observers will reach a mutually acceptable
agreement on what is going on when studying a causal process. In this
framework, a larger class of observers, beyond those in Newtonian
and Special Relativity frameworks, can connect their laboratory time
duration of a process to the proper-time duration measured by an observer
in a co-moving frame along the time-like process. The essential ingredient
of GR is the invariance of the proper-time interval $d\tau$ and the
proper-length interval $dl$; this is achieved by the notion of 
parallel transport 
that preserves the magnitude
of a vector upon its transport to nearby points in the configuration
space-time. The symmetry transformations of the space-time associated
with this larger class of observers are the largest possible set the 
diffeomorphisms of the space-time coordinates.
A theory that has such symmetry is called covariant theory.
All modern successful theories in physics are build to be explicitly
covariant \cite{2010CQGra..27u5018G,VGG2002Varna}.

Considering the above view of describing physical reality, and in
particular, that any physically acceptable observer can use their
own coordinate time as parametrization for a physical process then
it seems reasonable to impose 
the principle of reparametrization invariance 
along with the principle of the covariant formulation 
when constructing models of the physical processes \cite{VGG2002Varna,VGG2002Kiten,VGG2003}.
This means that along with the laboratory coordinates that label the
events in the local space-time of an observer, who is an arbitrary
and therefore can choose the coordinates in any way suitable, 
for the description of a natural phenomenon within the means of the laboratory
apparatus. The observer should also be free to choose an arbitrary
parametrization of the process as long as it is useful for the process
considered. As long as the formulation of the model is covariant then
there would be a suitable diffeomorphism transformation between any
two physical observers that will allow them to reach agreement on
the conclusions drawn from the data. Thus the process is independent
of the observer's coordinate frame. 
The reparametrization invariance of the process 
then means that the process is also independent, not
only on the coordinate frame of the observer, but it is also independent
on the particular choice of process parametrization selected by the
observer who is studying the process. Formulating a covariant theory
is well known in various sub-fields of physics, but if one embraces
\textit{the principle of reparametrization invariance} then there
are at least three important questions to be addressed: 
\begin{enumerate}\vspace{-3pt}
\item How do we construct such models? \vspace{-6pt}
\item What is the mathematical framework and what are the implications of such models?\vspace{-6pt}
\item What is the meaning/role of an arbitrarily time-parameter
$\lambda$ for a particular~process? \vspace{-6pt}
\end{enumerate}

The first question, ``How do we construct reparametrization invariant models?''
has been already discussed, in general terms, by the authors in a previous
publications \cite{VGG2002Kiten,VGG2002Varna,sym13030379} along with further
relevant discussions of the possible relations to other modern theories
and models \cite{VGG2002Varna,VGG2003,VGG2005}. The important line
of reasoning is that fiber bundles 
provide the mathematical framework for classical mechanics, field
theory, and even quantum mechanics when viewed as a classical field
theory. Parallel transport, covariant differentiation, and 
gauge symmetry 
are very important structures associated with fiber bundles
\cite{VGG2002Kiten,Pauli1958}.
When asking: ``What structures are important to physics?'', one should
also ask: ``Why one fiber bundle should be more `physical' than another?'',
``Why does the `physical' base manifold seems to be a four-dimensional
Minkowski manifold?'' \cite{VGG2002Kiten,VGG2002Varna,Borstnik2000,vanDam2001,Sachoglu2001},
and ``How should one construct an action integral for a given fiber bundle?''
\cite{VGG2002Varna,Kilmister1967,Carinena1995,Feynman1965,Gerjuoy1983,Rivas2001}.
Starting with the tangent or cotangent bundle seems natural because
these bundles are related to the notion of classical point-like objects.
Since we accumulate and test our knowledge via experiments that involve
classical apparatus, the physically accessible fields should be generated
by matter and should couple to matter as well. Therefore, the 
matter Lagrangian should contain the interaction fields, not their derivatives, 
with which classical matter interacts
\cite{Dirac1958.333D,VGG2002Kiten,VGG2005}.
The important point here is that probing and understanding physical
reality goes through a classical interface that shapes our thoughts
as classical causality chains. Therefore, understanding the essential
mathematical constructions in classical mechanics and classical field
theory is important, even though quantum mechanics and quantum field
theory are regarded as more fundamental than their classical counterparts.

In particular, the results relevant to Question 1 and 2 above seems to justify the existence only of
electromagnetic and gravitational interactions, as we know them, at a classical level within the
Lagrangian framework \cite{sym13030379}.

Two approaches, the Hamiltonian 
and the Lagrangian framework, 
are very useful in physics 
\cite{Kilmister1967,Goldstain1980,Gracia2001,Carinena1995,Deriglazov2011,Deriglazov2016,Nikitin-StringTheory}.
In general, there is a transformation that relates these two approaches.
For a reparametrization-invariant theory~\cite{sym13030379,Deriglazov2011,Deriglazov2016,1996PhRvD..53.7336L,2016CQGra..33f5004G,Kleinert1989},
however, there are problems in changing from Lagrangian to the Hamiltonian approach
\cite{Goldstain1980,Deriglazov2016,Deriglazov2011,Gracia2001,Rund1966,Lanczos1970,Nikitin-StringTheory}.

Given the remarkable results in \citep{sym13030379} due to the idea of reparametrization invariance, 
it is natural to push the paradigm further and to address  point 2 above, and to seek a suitable  
Hamiltonian formulation along with a relevant quantum framework. Some of the problems faced by 
reparametrization invariant systems studied here are also relevant to string theory and general relativity.
In this respect the lessons learned could be relevant to the understanding of space, time, 
and the quantum phenomenon \cite{STQEmergenceBook}.

In this paper, the problems related to 
changing from Lagrangian to the Hamiltonian approach are illustrated and their resolution for the
simplest one-dimensional reparametrization-invariant systems relevant to the physical reality 
as well as in the case of the relativistic particle in any dimension are discussed. 

The relativistic particle Lagrangian is used to justify the importance
of reparametrization-invariant systems and in particular the first-order
homogeneous Lagrangians in the velocities. The usual gravitational
interaction term along with the observational fact of finite propagational
speed is used to justify the Minkowski space-time physical reality.
The justification implies only one time-like coordinate in addition
to the spatial coordinates along which particles propagate with a
finite speed. By using the freedom of choosing time-like parametrization
for a process, it is argued that the corresponding causal structure
results in the observed common Arrow of Time and non-negative masses
for the physical particles. The meaning of the time parameter $\lambda$
is further investigated within the framework of reparametrization-invariant
systems. Such systems are studied from the point of view of the Lagrangian
and extended Hamiltonian formalism. The extended Hamiltonian formulation
is using an extended Poisson bracket $\left\llbracket ,\right\rrbracket $
which is generally covariant and applicable to reparametrization-invariant
systems. The extended Poisson bracket $\left\llbracket ,\right\rrbracket $
is defined over the extended phase-space (phase-space-time) and 
includes the coordinate time $t$ and the energy $p_{0}$ in a way consistent with 
the Canonical Quantization formalism. 
The corresponding extended Hamiltonian $\boldsymbol{H}$
defines the classical phase space-time of the system via the Hamiltonian
constraint $\boldsymbol{H}=0$ and guarantees that the Classical Hamiltonian
$H$ corresponds to $c p_{0}$ the energy of the particle
when the parametrization $\lambda=ct$ is chosen. Furthermore, if the
extended Hamiltonian for a classical system is quantized 
($\boldsymbol{H}\rightarrow\boldsymbol{\hat{H}}$)
by following the Canonical Quantization formalism and the corresponding
Hilbert space $\mathcal{H}$ is defined via the extended Hamiltonian
$\boldsymbol{\hat{H}}\Psi=0$ then the Schr\"odinger's equation emerges
naturally and the principle of superposition of quantum states is justified. 
A connection is demonstrated between the positivity
of the energy $E=cp_0>0$ and the normalizability of the wave function
by using the extended Hamiltonian that is relevant for the proper-time
parametrization. It is demonstrated that the choice of the extended
Hamiltonian $\boldsymbol{H}$ is closely related to the meaning of
the process parameter $\lambda$. The two familiar roles that $\lambda$
can take upon the coordinate time $t$ and the proper-time $\tau$ 
are illustrated using the simplest one-dimensional reparametrization
invariant systems. In general, $\lambda$ can also be the proper length
along the path of a particle for appropriately chosen transformation
generator similar to $\boldsymbol{H}$. 

The formalism is further illustrated in more details for the case of the relativistic particle in  Appendix \ref{Appendix} section of the paper.

The discussion starts by first reviewing the main points from 
\cite{VGG2002Kiten,VGG2002Varna,VGG2003,sym13030379}
as pertained to point particles: Section \ref{sec:Justifying-the-Reparametrization-Invarience}
has a Section 
 \ref{subsec:Relativistic-Particle-Lagrangian} on
the Lagrangian for a relativistic particle as an example of a reparametrization-invariant system, 
followed by a Section \ref{subsec:Homogeneous-Lagrangians-of-first-order}
where the general properties of homogeneous Lagrangians in the velocities
are stated, the section concludes with a list of pros and cons of
the first-order homogeneous Lagrangians. The next Section \ref{sec:One-Time-Physics Only}
is revisiting the argument why a space-time with a maximum speed of
propagation through space, when modeled via first-order homogeneous
Lagrangian based on a metric tensor, should be locally a Minkowski
space-time with common Arrow of Time and non-negative mass for the
particles. Section \ref{sec:From-L-to-H} is a brief review
of Lagrangian and Hamiltonian Mechanics and the problem of the Hamiltonian
constraint $H=0$ for systems based on first-order homogeneous Lagrangians.
In Section \ref{sec:From-Classical-to-QM}, the Canonical Quantization
is used as justification for the introduction of the extended covariant
Hamiltonian framework within which the Hamiltonian constraint $\boldsymbol{H}=0$
can be used to define the phase space-time, as well as to justify
the Schr\"odinger's equation as a consequence of applying Canonical
Quantization to the extended Hamiltonian framework. The meaning of
the process parameter $\lambda$ within the extended Hamiltonian framework
is discussed in Section 
 \ref{sec:The-Meaning_of-the-process-parameter} using
the simplest possible one-dimensional reparametrization invariant systems. 
The conclusions and discussions are given in Section \ref{sec:Conclusions-and-Discussions}.
The formalism is illustrated in more details for the case of the relativistic particle in the Appendix \ref{Appendix}. 

\section{Justifying the Reparametrization Invariance (RI) \label{sec:Justifying-the-Reparametrization-Invarience}}

\subsection{Relativistic Particle Lagrangian \label{subsec:Relativistic-Particle-Lagrangian}}

From everyday experience, we know that localized particles move with
a finite speed in a three-dimensional space. However, in an extended-configuration space
(space-time), when the time is added as a coordinate ($x^{0}=ct$), massive particles
move along a space-time trajectory such that $u\cdot u=1$. 
Here, $u^{\mu}$ are the coordinates of a general 4-velocity vector 
\mbox{$v^{\mu}=dx^{\mu}/d\lambda$} but with a special choice of the parametrization parameter $\lambda$;
that is,\linebreak  $u^{\mu}=dx^{\mu}/d\tau$. While $\lambda$ is an arbitrary parametrization, 
$\tau$ is a special choice of parametrization that is invariant with respect to 
any coordinate transformations between reasonable physical observers, it is the proper-time
($\tau$) mathematically defined via a metric tensor $g_{\mu\nu}$ ($d\tau^{2}=g_{\mu\nu}dx^{\mu}dx^{v}$). 
In particular when the metric tensor takes the form of the Minkowski metric $(g_{\mu\nu}=\eta_{\mu\nu})$ 
then one can talk about local Lorentz equivalent observers. 
In this case, the action for a massive relativistic particle
has a nice geometrical meaning:
the ``time distance'' along the particle trajectory \cite{Pauli1958}:
\begin{eqnarray}
&& S_{1} =  \int d\lambda L_{1}(x,v) =  \int d\lambda\sqrt{g_{\mu\nu}v^{\mu}v^{\nu}},\label{S1}\\
&& \sqrt{g_{\mu\nu}v^{\mu}v^{\nu}} \rightarrow \sqrt{g_{\mu\nu}u^{\mu}u^{\nu}}=1 
\Rightarrow S_{1} \rightarrow \int d\tau.\nonumber 
\end{eqnarray}
However, for massless particles, such as photons, the
4-velocity is a null vector ($g_{\mu\nu}v^{\mu}v^{\nu}=0$). 
Thus, proper time is not well defined and furthermore, one has to
use a different Lagrangian to avoid problems due to division by zero
when evaluating the final Euler--Lagrange equations. The appropriate
``good'' action is then \cite{Pauli1958}: 
\begin{equation}
S_{2}=\int  L_{2}(x,v) d\lambda=\int  g_{\mu\nu}v^{\mu}v^{\nu}d\lambda. \label{S2}
\end{equation}
For a massive particle, the Euler--Lagrange equations obtained from $S_{1}$ and
$S_{2}$ are equivalent, this equivalence is discussed in more detail later.
In the above discussion, it has been considered an arbitrary parametrization $\lambda$ 
and the proper-time parametrization $\tau$ for a massive particle. The physical meaning of the 
proper-time $\tau$ is usually considered to be the passing of clock time of 
a co-moving observer. Another important parametrization is the coordinate time $t$
corresponding to the clock time of an arbitrary physical observer that is studying 
the motion of the massive particle. Contemporary physics models are expected to be 
invariant with respect to coordinate transformations between physical observers.
This is achieved by constructing Lagrangians as a scalar object from various vector 
and  tensor quantities that correspond to the measurements of an arbitrary observer. 
In mathematical terms, this is a diffeomorphism invariance of the coordinate space. Thus, the physics content of
a process under study is the same and therefore independent of the  coordinate system of an observer.
Clearly, diffeomorphism invariance is an important symmetry that reflects the expectation 
that observing a process should not affect the process itself. Thus, various observers 
should find a way to understand each-others measurements in a consistent way  as long as they 
pertain to the same process under the study. 

In the example above, the process under study is the motion of a massive particle. 
In this respect, the process corresponds to a one-dimensional manifold that is 
a curve in a higher dimensional space-time. It seems that the relationships of the points along the curve, 
in particular, the ordering of the points and their relative measures, 
should be something about the curve (the process under study). 
Thus, various observers should be able to find a consistent way to understand the curve and its properties. 
Therefore, the description of the curve should be independent of the choices an observer can make 
in order to describe the curve. In particular, the choice of parametrization of the curve 
should be irrelevant to the understanding of the corresponding process. 
Thus, a reparametrization invariant formulation
would be the corresponding symmetry that the description should obey. While a model built on 
$L_{1}$ above does obey such symmetry,  its quadratic version based on $L_2$ does not seem 
to obey it, even though the  corresponding Euler--Lagrange equations are equivalent. 
Even more, the Euler--Lagrange equation does obey parametrization-rescaling symmetry that is easily seen 
when the Lagrangian is a homogeneous function of the velocity (see below).
A way to resolve this puzzle is to recognize that $L_2$ can be viewed as 
a reparametrization invariant Lagrangian in a particular fixed gauge \cite{Green1987book}:
\begin{equation}
S^\prime_{2}=\int  g_{\mu\nu}v^{\mu}v^{\nu} e^{-1} d\lambda. \label{S'2}
\end{equation}
Here $e$ is an auxiliary field that makes the action $S^\prime_{2}$ reparametrization invariant by choosing 
$e \rightarrow \tilde{e}=e \, d\lambda/d\tilde{\lambda}$ when $\lambda \rightarrow  \tilde{\lambda}$. 
Since now  $S^\prime_{2}$ is reparametrization invariant then one can choose 
a gauge parametrization such that $e=1$ and thus arriving at $S_{2}$ but under 
proper-time parametrization $\lambda=\tau$ where $g_{\alpha\beta}u^{\alpha}u^{\beta}=1$.
Having to choose the gauge such that $g_{\alpha\beta}v^{\alpha}v^{\beta}$ is constant guarantees
the equivalence of $L_1$ and $L_2$ \cite{Rizzuti201905,sym13030379}. 

\subsection{Equivalence of Homogeneous Lagrangians}

The above equivalence of $S_1$ and $S_2$ 
could be demonstrated on a more complicated Lagrangian as 
a specific choice of parametrization such that 
$g_{\alpha\beta}\left(x\right)v^{\alpha}v^{\beta}$ is constant \cite{Rizzuti201905,sym13030379}.
Indeed, if one starts with the reparametrization invariant Lagrangian:

\begin{equation}
L=qA_{\alpha}v^{\alpha}-m\sqrt{g_{\alpha\beta}(x)v^{\alpha}v^{\beta}}\label{RI-L4EM}
\end{equation}
and defines proper-time 
gauge $\tau$ such that: 
\begin{equation}
d\tau=\sqrt{g_{\alpha\beta}dx^{\alpha}dx^{\beta}}\Rightarrow\sqrt{g_{\alpha\beta}u^{\alpha}u^{\beta}}=u\cdot u=1,\label{uu=1}
\end{equation}
then one can effectively consider 
$$L=qA_{\alpha}u^{\alpha}-(m-\chi)\sqrt{g_{\alpha\beta}u^{\alpha}u^{\beta}}-\chi$$
as the model Lagrangian. Here $\chi$ is a Lagrange multiplier to
enforce $u\cdot u=1$ in (\ref{uu=1}) that breaks
the reparametrization invariance explicitly. 
Then one can write it as:
$$L=qA_{\alpha}u^{\alpha}-(m-\chi)\frac{g_{\alpha\beta}u^{\alpha}u^{\beta}}{\sqrt{g_{\alpha\beta}u^{\alpha}u^{\beta}}}-\chi$$
and using $u\cdot u=1$ in (\ref{uu=1}) one arrives at: 
\begin{equation}
L=qA_{\alpha}u^{\alpha}-(m-\chi)g_{\alpha\beta}u^{\alpha}u^{\beta}-\chi.\label{constrained-L4EM}
\end{equation}
One can deduce a specific value for $\chi$ ($\chi=m/2$) by requiring
that (\ref{RI-L4EM}) and (\ref{constrained-L4EM})
produce the same Euler--Lagrange equations under the constraint $u\cdot u=1$ in (\ref{uu=1}).
Then, by dropping the overall constant term, this finally results
in the familiar equivalent Lagrangian: 
\begin{equation}
L=qA_{\alpha}u^{\alpha}-\frac{m}{2}g_{\alpha\beta}u^{\alpha}u^{\beta}.\label{quadratic-L4EM}
\end{equation}
where $\tau$ has the usual meaning of proper-time parametrization such that 
$u\cdot u=1$ in (\ref{uu=1}), 
but it is imposed after deriving all the equations from the Lagrangian under consideration.

The equivalence between $S_{1}$ and $S_{2}$ is very robust. Since
$L_{2}$ is a homogeneous function of order $2$ with respect to the
four-velocity $\vec{v}$, the corresponding Hamiltonian function ($H=v\partial L/\partial v-L$)
is exactly equal to $L$ ($H(x,v)=L(x,v)$). Thus, $L_{2}$ is conserved,
and so is the length of $\vec{v}$ and therefore $L_{1}$ is conserved as well.
Any homogeneous Lagrangian in
$\vec{v}$ of order $n\neq1$ is conserved because $H=(n-1)L$. 
When $dL/d\lambda=0$, then one can show that the Euler--Lagrange equations
for $L$ and $\tilde{L}=f\left(L\right)$ are equivalent under certain
minor restrictions on $f$ \cite{sym13030379}. 

To see this, consider the Euler--Lagrange equation for $L$: 
\[
\frac{d}{d\lambda }\left( \frac{\partial L}{\partial v^{i}}\right) -
\frac{\partial L}{\partial x^{i}}=0 ,
\]
and compare it with the Euler--Lagrange equation for $\tilde{L}=f\left(L\right)$ 
that can be written as:
\[
\left( \frac{f^{\prime \prime }}{f^{\prime }}\frac{dL}{d\lambda }\right) \left( 
\frac{\partial L}{\partial v^{i}}\right) +\frac{d}{d\lambda }\left( \frac{
\partial L}{\partial v^{i}}\right) -\left( \frac{\partial L}{\partial x^{i}}\right) =0 
\]
Clearly, these equations will be equivalent if the Lagrangians 
$L$ and $\tilde{L}=f\left(L\right)$  are constants of the motion; that is, $d L/d\lambda=0$,
and $f^{\prime }$ and $f^{\prime\prime }$ are well behaved.

This is an interesting type of 
equivalence that applies to homogeneous Lagrangians \linebreak $(L(\beta v)=\beta^{n}L(v)).$
It is different from 
the usual equivalence $L\rightarrow\tilde{L}=L+d\Lambda/d\lambda$ or the more 
general equivalence discussed in Reference  \cite{Hojman1981,Rivas2001,Baker2001,Gracia2001}.
Any solution of the Euler--Lagrange equation for $\tilde{L}=L^{\alpha}$
would conserve $L=L_{1}$ since $\tilde{H}=(\alpha-1)L^{\alpha}$ when $\alpha\neq1$,
while for  $\alpha=1$ it can be enforced as a choice of parametrization.
For example, as demonstrated above, one can always make the choice of proper-time parametrization 
$\lambda=\tau$ for a massive particle. 
All these solutions are solutions of the Euler--Lagrange equation for
$L$ as well; thus, $L^{\alpha}\subset L$. The fact that the models based on $L_2$ are only a subset of $L_1$
implies that $L_1$ has a special role due to its richer applicability to physical systems. 
In particular, such Lagrangians are of unique type relevant to the Weyl Integrable geometry \cite{Bouvier197804}.
Weyl's Integrable geometry provides a framework that is likely to be relevant to physics \cite{Maeder197903} and
may have a far reaching consequences  for cosmology \cite{Maeder201811}.
In general, conservation of $L_{1}$ is not guaranteed since $L_{1}\rightarrow L_{1}+d\Lambda/d\lambda$
is also a homogeneous Lagrangian of order one equivalent to $L_{1}$.
This suggests that there may be a choice of $\lambda$, a ``gauge
fixing'', such that $L_{1}+d\Lambda/d\lambda$ is conserved even if
$L_{1}$ is not. The above discussion applies to any homogeneous Lagrangian
that has no explicit time dependence.

\subsection{Homogeneous Lagrangians of First Order \label{subsec:Homogeneous-Lagrangians-of-first-order}}

Suppose we know nothing about classical physics, which is mainly
concerned with trajectories of point particles in some space $M$,
but we are told we can derive it from a variational principle if we
use the right action integral $S=\int Ld\lambda$. By following the above
example we wonder: ``should the smallest `distance' be the guiding
principle?'' when constructing $L$. If yes, ``How should it be defined
for other field theories?'' It seems that a reparametrization-invariant
theory can provide us with a 
metric-like structure \cite{Rund1966,Lanczos1970,Landau-Lifshitz1975,Gerjuoy1983},
and thus a possible link between field models and geometric models
\cite{Rucker1977}.

In the example of the relativistic particle, the Lagrangian and the
trajectory parameterization have a geometrical meaning. In general,
however, parameterization of a trajectory is quite arbitrary for any
observer. If there is such thing as the smallest time interval that
sets a space-time scale, then this would imply a 
discrete space-time structure since there may not be any events in the smallest time interval. 
The Planck scale is often considered
to be such a special scale \cite{Magueijo2002,2003ApJ...587L...1R}.
Leaving aside hints for quantum space-time from loop quantum gravity
and other theories, we ask: ``Should there be any preferred trajectory
parameterization in a smooth 4-dimensional space-time?'' and ``Are we not free
to choose the standard of distance (time, using natural units $c=1$)?''
If so, then one should have a smooth continuous manifold and
the theory should not depend on the choice of parameterization. 

If one examines the Euler--Lagrange equations carefully: 
\begin{equation}
\frac{d}{d\lambda}\left(\frac{\partial L}{\partial v^{\alpha}}\right)=\frac{\partial L}{\partial x^{\alpha}},\label{Euler--Lagrange equations}
\end{equation}
one will notice that any homogeneous Lagrangian of order $n$ ($L(x,\alpha\vec{v})=\alpha^{n}L(x,\vec{v})$) provides a 
reparametrization invariance of the equations under the rescaling transformations of the parametrization
$\lambda\rightarrow\lambda/\alpha,\vec{v}\rightarrow\alpha\vec{v}$.
Next, note that the action $S$ involves an integration that is a
natural structure for orientable manifolds ($M$) with an $n$-form
of the volume. Since a trajectory is a one-dimensional object, then
what is one looking at is an embedding\linebreak $\phi:\Bbb{R}^{1}\rightarrow M$.
This means that one is pushing forward the tangential space $\phi_{*}:T(\Bbb{R}^{1})=\Bbb{R}^{1}\rightarrow T(M)$,
and is pulling back the cotangent space $\phi^{*}:T(\Bbb{R}^{1})=\Bbb{R}^{1}\leftarrow T^{*}(M)$.
Thus, a 1-form $\omega$ on $M$ that is in $T^{*}(M)$ ($\omega=A_{\mu}\left(x\right)dx^{\mu}$)
will be pulled back on $\Bbb{R}^{1}$ ($\phi^{*}(\omega)$) and there
it should be proportional to the volume form on $\Bbb{R}^{1}$ ($\phi^{*}(\omega)=
A_{\mu}\left(x\right)(dx^{\mu}/d\lambda)d\lambda\sim d\lambda$),
allowing it to be integrated $\int\phi^{*}(\omega)$: 
\[
\int\phi^{*}(\omega)=\int Ld\lambda=\int A_{\mu}\left(x\right)v^{\mu}d\lambda.
\]

Therefore, by selecting a 1-form $\omega=A_{\mu}\left(x\right)dx^{\mu}$
on $M$ and using $L=A_{\mu}\left(x\right)v^{\mu}$ one is actually
solving for the embedding $\phi:\Bbb{R}^{1}\rightarrow M$ using a
chart on $M$ with coordinates $x:M\rightarrow\Bbb{R}^{n}$. The Lagrangian
obtained this way is homogeneous of first-order in $v$ with a very
simple dynamics. The corresponding Euler--Lagrange equation is $F_{\nu\mu}v^{\mu}=0$
where $F$ is a 2-form ($F=dA$); in electrodynamics, this is the
Faraday's tensor. If one relaxes the assumption that $L$ is a pulled
back 1-form and assume that it is just a 
homogeneous Lagrangian of order one, then one may find a reparametrization-invariant
theory that could have an interesting dynamics. The above mathematical
reasoning can be viewed as 
justification for the known classical forces of electromagnetism and gravitation and perhaps even of 
new classical fields beyond electromagnetism and gravitation
\cite{VGG2002Kiten,VGG2002Varna,VGG2005, sym13030379}.

\subsection{Pros and Cons of Homogeneous Lagrangians of the First Order\label{subsec:Pros-and-Cons}}

Although most of the features listed below are more or less self-evident, 
it is important to compile a list of properties of the first-order
homogeneous Lagrangians in the velocity $\vec{v}$.

\noindent
Some of the good properties of a theory with a first-order homogeneous Lagrangian are \cite{sym13030379}: 
\begin{itemize}\vspace{-3pt}
\item[(1)] First of all, the action $S=\int L(x,\frac{dx}{d\lambda})d\lambda$ is
a reparametrization invariant. \vspace{-6pt}
\item[(2)] For any Lagrangian $L(x,v=\frac{dx}{dt})$ one can construct a 
reparametrization-invariant Lagrangian by enlarging the configuration space $\{x\}$ to
an extended configuration space---the space-time $\{ct,x\}$ \cite{Goldstain1980, Deriglazov2011, Deriglazov2016}. 
However, it is an open question whether there is a full equivalence
of the corresponding Euler--Lagrange equations. \vspace{-6pt}
\item[(3)] Parameterization-independent path-integral quantization could be
possible since the action $S$ is reparametrization invariant \cite{Fradkin1991}. \vspace{-6pt}
\item[(4)] The reparametrization invariance may help in dealing with singularities
\cite{Kleinert1989}. \vspace{-6pt}
\item[(5)] It is easily generalized to $D$-dimensional extended
objects ($p$-branes /$d$-branes) \cite{VGG2002Kiten,VGG2002Varna}. \vspace{-3pt}
\end{itemize}
The list of trouble-making properties in a theory with a first-order
homogeneous Lagrangian includes: 
\begin{itemize}\vspace{-3pt}
\item[(1)] \textls[-12]{There are constraints among the Euler--Lagrange equations \cite{Goldstain1980, Deriglazov2016},
since $\det\left(\frac{\partial^{2}L}{\partial v^{\alpha}\partial v^{\beta}}\right)=0$.} \vspace{-6pt}
\item[(2)] It follows that the Legendre transformation ($T\left(M\right)\leftrightarrow T^{*}\left(M\right)$),
which exchanges velocity and momentum coordinates $(x,v)\leftrightarrow(x,p)$,
is problematic \cite{Gracia2001}. \vspace{-6pt}
\item[(3)] There is a problem with the canonical quantization approach since
the Hamiltonian function is identically ZERO ($H\equiv0$) \cite{Lanczos1970,Nikitin-StringTheory},\vspace{-3pt}
\end{itemize}

The pro (2)  and the con (3) above are of key importance. 
The procedure that can be utilized as mentioned in pro (2) above is very simple:
$L(x,\frac{dx}{dt})\rightarrow {\dot{t}}L(x,\frac{\dot{x}}{\dot{t}})$ 
where the dotted notation is a derivative with respect to the parametrization $\lambda$,
that is, ${\dot{t}}=\frac{dt}{d\lambda}$ and ${\dot{x}}=\frac{dx}{d\lambda}$ 
\cite{Deriglazov2011,Deriglazov2016}. 
This means that every Lagrangian based theory can be reformulated in a
reparametrization invariant form. 
This is a different symmetry from the diffeomorphism invariance of the theory, 
which is still satisfied by  construction. However, the parameter $\lambda$ 
does not have to be the typical physical time parameterization of a process 
like its own process time---the proper time $\tau$, 
nor the coordinate time $t$ for the observer that is studying the process.
In this sense, $\lambda$ could be truly arbitrary and thus demonstrating the 
existence of a larger class of theories that do satisfy  
the principle of reparametrization invariance as discussed in the introduction.
The problem with these larger class of theories is in the con (3). 
Which makes the standard quantization treatment quite difficult and unusual due to
the presence of constraints among the equations of motion con (1) above \cite{Deriglazov2011}.
In this paper, we are mostly concerned with physical processes that 
are associated with one-dimensional manifolds and their reparametrization.
However, as the pro (5) suggests, the formulation is relevant to two-dimensional  
sub-manifolds, which is the domain of string theory, and extends to
high-dimensional sub-manifolds with reparametrization invariant Lagrangians 
such as the Nambu--Gotto Lagrangians \cite{VGG2005,Nikitin-StringTheory}.
The reparametrization invariance, 
which is a diffeomorphism of the submanifold corresponding to a physical process, 
is a far-reaching idea and it is different from 
the coordinate diffeomorphisms of the target space~\cite{sym13030379}. 
However, it is beyond the scope of this paper to go into string-theory, 
p-branes, and gravity that represent sub-manifolds with dimension bigger than one.

Constraints among the equations of motion are not an insurmountable problem since
there are procedures for quantizing such theories 
\cite{Dirac1958.326D,Teitelboim1982,Henneaux1992ig,Sundermeyer1982,deLeon2013,Nikitin-StringTheory}.
For example, instead of using $H\equiv0$ one can use some of the
constraint equations available, or a conserved quantity, as Hamiltonian
for the quantization procedure \cite{Nikitin-StringTheory}.
Changing coordinates $(x,v)\leftrightarrow(x,p)$ seems to be difficult,
but it may be resolved in some special cases by using the assumption
that a gauge $\lambda$ has been chosen so that 
$L\rightarrow L+\frac{d\Lambda}{d\lambda}=\tilde{L}=const$.
The above-mentioned quantization troubles will not be discussed here since
they are outside of the scope of this paper. A new approach that resolves
$H\equiv0$ and naturally leads to a Dirac-like equation is under
investigation, for some preliminary details see Reference  \cite{VGG2002Varna}.
Here, the focus is on the understanding of the meaning and the role of the general
parameter $\lambda$ by extending the Hamiltonian framework to an
extended phase-space (phase-space-time) with a \textit{covariant extended
Poisson Bracket }$\left\llbracket ,\right\rrbracket $, which is consistent
with the \textit{Canonical Quantization process}, along with an \textit{extended
Hamiltonian} $\boldsymbol{H}$ that defines the \textit{extended phase-space-time}
via $\boldsymbol{H}\equiv0$. By following the Canonical Quantization formalism 
($\boldsymbol{H}\rightarrow\boldsymbol{\hat{H}}$) 
the Hilbert space of the quantum system can be
defined via the corresponding  extended Hamiltonian  $\boldsymbol{\hat{H}}$
as the linear space of states $\Psi$ that satisfy $\boldsymbol{\hat{H}}\Psi=0$. 
As a byproduct, one can use this formulation
to justify the Schr\"odinger's equation. A similar approach to the Schr\"odinger's equation 
has been discussed in Reference  \cite{Deriglazov2009,Deriglazov2011,Deriglazov2016}. 
Furthermore, we do not concern ourselves with the questions about the algebra of observables 
nor with issues of unboundedness of important physical operators 
or alternative approaches to the quantization framework. Such issues are outside of the scope of the current paper.
Finally, before we discuss the extended Hamiltonian framework, we would like to use the structure
of the first-order homogeneous Lagrangians to mathematically
justify few other important features of the physical reality that we often take for granted.

\section{One-Time Physics, Causality, Arrow of Time, and the Maximum Speed of~Propagation}
\label{sec:One-Time-Physics Only}

In our everyday life, most of us take time for granted, but there
are people who are questioning its 
actual existence or consider 
models with more than one time-like coordinate
\cite{1983PhRvD..27.2885P,1986FoPh...16..437G,1994GReGr..26.1267G,2002PhRvD..66d4020E,
2006tgrq.conf...79E,Albrecht2008,2010arXiv1010.2968B,2011arXiv1110.3296V,2011CQGra..28c5006B,
2011JPhCS.306a2013P,2012FoPh...42.1384W,2013GReGr..45..911B,2015FoPh...45..691V,
2017PhRvD..95d3510M,Anderson2017,Borstnik2000,Elci2010,Renner2017,
vanDam2001,2012PhyEs..25..403V,2009RSPSA.465.3023C}.
Since we are trying to understand the meaning of an arbitrary time-like
parameter $\lambda$ within the framework of reparametrization invariant
systems, it seems important to think about the possible number of
time-parameters. Here, we briefly argue that a one-time-physics, in
case of massive point particles, is essential to assure 
causality via finite propagational speed
through space for such massive point particles \cite{VGG2002Varna}.
Then the 
common arrow of time, which is often viewed as related to the 
increasing entropy as commanded by second law of thermodynamics
\cite{2004hep.th...10270C,2005GReGr..37.1671C,2013EPJWC..5802005G,1983Natur.304...39P,
1984Natur.312..524D,2011arXiv1110.3296V,2011PhRvD..83d3503L}, 
becomes instead a consequence of the 
positivity of the rest mass.

\subsection{One-Time Physics, Maximum Speed of Propagation, and the Space-Time Metric Signature}
\label{Space-Time Metric Signature}

Why the space-time seems to be one time plus three spatial dimensions
have been discussed by using arguments a la Wigner \cite{Borstnik2000,vanDam2001}.
However, these arguments are deducing that the space-time is $1+3$ because
only this signature is consistent with particles with finite spin.
However, one should turn this argument backwards claiming that
one should observe only particles with finite spin because the signature is $1+3$. 
A thermodynamic selection principle of the ($1+3$) nature of the universe 
has been recently discussed \cite{2016EL....11340006G}. 
And yet, there is an alternative argument for the emergence of the 
apparent Lorentzian dynamics of the usual field theories due to a scalar clock field 
that is playing the role of the physical time \cite{2013PhRvD..87f5020M}.
Furthermore, there is an argument for ($1+3$) nature of the universe based on 
four fundamental principles of physics namely: Causality, General Covariance, 
Gauge Invariance, and Renormalizability \cite{2013arXiv1303.5634P}. 
This can be taken even further to dynamically generate a fifth dimension 
\cite{2001PhRvL..86.4757A}, which contains an extra special dimension in contrast 
to the traditional arguments for three special dimensions  
\cite{1997CQGra..14L..69T,2005SHPMP..36..113C}.

\newpage
Here we revisit the argument that only one-time physics is
consistent with a finite spatial propagational speed \cite{VGG2002Varna}.
The local Lorentz symmetry 
implies the existence of a local observer with Minkowski like coordinate frame.\\

The main assumptions are: 
\begin{itemize}\vspace{-3pt}
\item[I.]Gravity-like term $\sqrt{g(\vec{v},\vec{v})}$ is always present in the matter Lagrangian.\vspace{-6pt}
\item[II.] The corresponding matter Lagrangian is a real-valued function. \vspace{-3pt}
\end{itemize}

This way, physical processes, like propagation of a particle, must be related to positive-valued
term $g(\vec{v},\vec{v})\geq0$. Here $\vec{v}$ is the rate of change
of the space-time coordinates with respect to some arbitrarily chosen
parameter $\lambda$ that describes the evolution of the process (propagation
of a particle). That is, $v^{\alpha}=dx^{\alpha}/d\lambda$. By speed
one should mean the magnitude of the spatial velocity with respect to a laboratory
time coordinate $x^{0}$  where $v_{space}^{i}=dx^{i}/dx^{0}$.

The use of a covariant formulation allows one to select a local coordinate
system so that the metric is diagonal $(+,+,..,+,-,...-)$. 
If one denotes the (+) coordinates as time coordinates and the ($-$) as spatial
coordinates, then there are three essential cases: 

\begin{itemize}
\item[(1)] {No time coordinates will contradict $g(\vec{v},\vec{v})\geq0$:
$$g(\vec{v},\vec{v})=-\sum_{\alpha}\left(v^{\alpha}\right)^{2}<0.$$} 
\item[(2)] {Two or more time coordinates---unconstrained spacial velocity $\vec{v}_{space}$:
$$g(\vec{v},\vec{v})=(v^{0})^{2}+(v^{1})^{2}-\sum_{\alpha=2}^{n}\left(v^{\alpha}\right)^{2}\Rightarrow1+w^{2}\geq\vec{v}_{space}^{2}.$$} 
\item[(3)]{Only one time coordinate enforces finite spacial velocity $\vec{v}_{space}$:
$$g(\vec{v},\vec{v})=(v^{0})^{2}-\sum_{\alpha=1}^{n}\left(v^{\alpha}\right)^{2}\Rightarrow1\geq\vec{v}_{space}^{2}.$$}
\end{itemize}
Clearly, for two or more time coordinates, one does not have finite coordinate
velocity ($\vec{v}_{space}^{2}=(dl/dx^{0})^{2} \rightarrow \vec{v}^{2}/c^2$)
bound from above by the speed of light $c$.
For example, when the coordinate time ($x^{0}$) is chosen so that 
$x^{0}=ct\Rightarrow$ $v^{0}=1$
then along another time-like coordinate $x^{1}$ the speed will be
$w^{2}=(dx^{1}/(cdt))^{2}$, which could be anything in
magnitude unless $dx^{1}/(cdt)$ has an upper bound. 
Therefore, if $w$ is not zero then there would be processes that will
exhibit deviations from the observed maximal speed $c$ of propagation. 
Thus, only the space-time with only one time accounts
for a strict finite spatial speed of propagation, 
as observed, where the finite spatial
velocity is bounded from above by the speed of light $v<c$. 
Therefore, a causal structure! When going from one point
of space to another, it takes time and thus there is a 
natural causal structure \cite{Bekenstein1993,Renner2017}.
The details of the causal structure will depend on the interactions
that can take place when two objects are at the same point in the
space-time since there is a natural future-and-past cone in such~space-time.

\subsection{Causality, the Common Arrow of Time, and the Non-Negativity of the Mass}
\label{positive-mass}

In the previous section, we deduced that the space-time
metric has to reflect that there is only one coordinate time and the
rest of the coordinates should be spatial which is a requirement for
finite spatial speed of propagation that induces causality. If nature
is really reparametrization-invariant, then any observer studying
a process can use its own time coordinate $t$ or any other suitable
time-parameter $\lambda$, to label an unfolding process. However,
when comparing to other observers who study the same process, it will
be more advantageous to use a proper-time parametrization
$\tau$ which is usually related to an observer who is following/moving
along with the process (particle propagation in its co-moving frame).
To be able to study a process using any laboratory time-coordinate
$t$ and to deduce the process proper-time parametrization
for the purpose of comparing to other arbitrary observers would
then imply a reparametrization-invariant symmetry. 

The process has to be related to a massive system because actual observers
are also massive and cannot move as fast as light or other massless
particles due to the previously deduced $1+n$ signature of the metric
and non-negativity of $g(\vec{v},\vec{v})\ge0$.
As long as there is a term $m(v)\sqrt{g(\vec{v},\vec{v})}$ in the
Lagrangian $L$ and $m(0)\ne0$ then the relationship\linebreak 
$m(0)d\tau=m(v_{space})\sqrt{g(\vec{v},\vec{v})}d\lambda$
could be used to define proper-time $\tau$. 
Since $g(\vec{v},\vec{v})>0$ then one can consider the positive branch of the 
square root function, otherwise upon utilizing the reparametrization symmetry
one can consider $\lambda\rightarrow -\lambda $ that will correspond 
to the positive branch of the square root function when combined into 
$\sqrt{g(\vec{v},\vec{v})}d\lambda$.
Note that $m(v_{space})$ is a homogeneous function of zero order since 
it depends only on the dimensionless special velocity 
$v^{i}=dx^{i}/dx^{0}=(dx^{i}/d\lambda)/(dx^{0}/d\lambda)$
that is invariant under reparametrization.
If one considers the obvious choice $\lambda=t$
for any particular laboratory observer then this would imply that
time is going forward for the observer as well as for the process---as long as $m(v)\not=0$ for any physical value of $v$; thus, $m(v)$ and $m(0)$
have the same sign and therefore $m(v)/m(0)>0$.
This would be valid for any two observers that can study each other's motion
as well. Therefore, all observers that can study at least one common
process in nature will find a common arrow of time 
$dt/d\tau = m(0)/(m(v)\sqrt{g(\vec{v},\vec{v})})>0$. 
Such ``positive flow'' of time is important when considering the standard 
Lagrangian formulation for a relativistic particle and its reformulation as
a reparametrization invariant system \cite{Deriglazov2011}. 
The  ``positive flow'' of time assures the preservation of the  sign of the mass term.
Furthermore, by using the freedom of parametrization an observer may decide to use 
same process to define the scale of the time interval for the clocks. 
By choosing the parametrization to be the coordinate time with time intervals as
measured by proper time intervals for a process, $d\lambda=dt=d\tau$, 
then this gives us the relation of the rest mass to moving mass 
$m(0)=m(v)\sqrt{g(\vec{v},\vec{v})}$
as deduced in the theory of Special Relativity \cite{Pauli1958}. 

All processes and observers should have the same sign of their mass,
if not then one can envision a non-interacting pair that has opposite
sign of their rest masses ($m'(0)=-m''(0)$) moving in the same way
(with $\vec{v}$ the same and thus $g(\vec{v},\vec{v})$ the same as well) 
with respect to us; one expects that they will have the same proper-time; however, as a pair their
proper-time would not be accessible to us since their combined rest mass is zero ($m'(0)+m''(0)=0$).
This situation creates a proper-time paradox if the two particles can be observed separately during the 
motion of the pair. This could well be the case of annihilating particle and anti-particle pair 
since then the notion of a proper-time of a photon is not well defined, 
however, a sub-system pair has never been observed as photon sub-structure.
Given that all non-zero rest masses have to be of the same sign and the usual relationship
between rest mass and energy ($E=mc^2$), one can conclude that $m(v)>0$. \textls[-12]{Thus, the non-negativity
of the masses of particles and the positivity of the mass of physical observers. 
One will see later that the positivity of the energy is
related to the positivity of the norm of the corresponding quantum
system in its proper-time quantization within the extended Hamiltonian
formalism---see Section~\ref{subsec: Positivity of the Energy}.}

This is illustrated in more details in Appendix \ref{Appendix} for the case of the 
relativistic particle where the relativistic mass of a moving particle is related to the relativistic factor 
$\gamma$ and the rest mass of the particle $m(v)=\gamma(v) m(0)$.

In the light of the above discussion, the Common Arrow of Time is a result of the
positivity of the mass 
and has nothing to do with the entropy of a closed system and the
second law of thermodynamics
\cite{2004hep.th...10270C,2005GReGr..37.1671C,2013EPJWC..5802005G,1983Natur.304...39P,
1984Natur.312..524D,2011arXiv1110.3296V,2011PhRvD..83d3503L}.
 
To be more accurate one should point out that the positive correlation between 
forward-increasing time and the increasing entropy due to the second law of thermodynamics is not a cause--effect relation. 
Increasing entropy in general is just the general manifestation of forward-time flow. 
It is known that, locally, a system’s entropy can be reduced but this does not change the flow of time. 
In our discussion above, the Common Arrow of Time is due to common processes between observers, 
“entanglement with the environment,'' which acts as a clock (\cite{STQEmergenceBook}, Chapter 17),
which is forced to be synchronized due to the positivity of the mass. 
Thus, stopping and reversing the time flow for a macroscopic system is practically impossible, 
as it would have to overcome the second law of thermodynamics. 
However, for microscopic systems, that could be decoupled from the environment and the observer, 
the time-flow could be inverted, and therefore time-symmetric laws would be appropriate.
Furthermore, there is a connection between the positivity of the mass and 
the positivity of the temperature in thermodynamics (\cite{Stuckelberg'74}, Chapter 1).
This of course results, in general, in a decreasing of the temperature as entropy increases for a closed system,
but this is the manifestation of the general tendency of decay towards the ground state of a system,
which is not considered to be the cause of the Common Arrow of Time.

\section{From Lagrangian to Hamiltonian Mechanics \label{sec:From-L-to-H}}

In this section, we review the main relevant equations and concepts
that we need to better understand the meaning of the parameter
$\lambda$ in reparametrization-invariant systems and the particular
roles it can take. To keep the notation simple we write the equations
as for a one-dimensional system but the equations can easily be re-expressed
for $n$-dimensional systems as well.

\subsection{Lagrangian Mechanics}

Given a Lagrangian $\mathrm{\mathit{L(t,q,v)}}$, with velocity $v=\dot{q}=\frac{dq}{dt}$,
one can derive the 
Euler--Lagrange equations of motion by minimizing the corresponding
action $\mathcal{A}$ = $\int L(t,q,\dot{q})dt$ along trajectories $\{q(t):\delta A=0\}$ 
\cite{Rund1966,Lanczos1970,Gerjuoy1983,Pauli1958,Landau-Lifshitz1975}:

\begin{equation}
\frac{dp}{dt}=\frac{\partial L}{\partial q},\label{Euler--Lagrange equations-2}
\end{equation}
where the generalized momentum $p$ is given as: 
\begin{equation}
p\coloneqq\frac{\partial L}{\partial v},\label{eq:Generalized momentum}
\end{equation}
If one considers the Hamiltonian function $H(t,q,v)$:

\begin{equation}
H=pv-L(q,v)\label{eq:Hamiltonian function}
\end{equation}
then the Euler--Lagrange equations can be written as 
Hamilton's equations 
\cite{Carinena1995,deLeon2013,Rund1966}:
\begin{equation}
v\coloneqq\frac{dq}{dt}=\frac{\partial H}{\partial p},\label{eq:generalized velocity-1}
\end{equation}
\begin{equation}
\frac{dp}{dt}=-\frac{\partial H}{\partial q},\label{eq:Hamiltonian dynamics}
\end{equation}

\textls[+12]{The full power of the Hamiltonian framework is realized if one can solve 
\linebreak Equation~(\ref{eq:Generalized momentum})} for all the velocities $v=dq/dt$ as functions of $(t,q,p)$; 
then one can study the system using $H(t,q,p)$ over the phase-space coordinates $(q,p)$.
The dynamical equations of the Lagrangian framework  (\ref{Euler--Lagrange equations-2})
are then replaced by the dynamical equations of the Hamiltonian framework
(\ref{eq:Hamiltonian dynamics}).

The rate of change of an observable $f(t,q,v)$ that is a function
of time, position, and velocity is then given by: 
\begin{equation}
\frac{df}{dt}=\frac{\partial f}{\partial q}\left(\frac{dq}{dt}\right)+\frac{\partial f}{\partial v}\left(\frac{dv}{dt}\right)+\frac{\partial f}{\partial t}
\end{equation}

In a similar way the rate of change of an observable over the phase-space $f(t,q,p)$ that is a function
of time, position, and momentum is then given by: 
\begin{equation}
\frac{df}{dt}=\frac{\partial f}{\partial q}\left(\frac{dq}{dt}\right)+\frac{\partial f}{\partial p}\left(\frac{dp}{dt}\right)+\frac{\partial f}{\partial t}
\end{equation}

When applied to the Hamiltonian function $H$ for solutions that satisfy the equations above, one obtains: 
\begin{eqnarray*}
\frac{dH}{dt}&=&\frac{\partial H}{\partial q}\left(\frac{dq}{dt}\right)
+\frac{\partial H}{\partial p}\left(\frac{dp}{dt}\right)+\frac{\partial H}{\partial t} =\\
&=&-\frac{\partial L}{\partial q}v+v\frac{dp}{dt}-\frac{\partial L}{\partial t} =\\
&=&-\frac{dp}{dt}v+v\frac{dp}{dt}-\frac{\partial L}{\partial t}=-\frac{\partial L}{\partial t}
\end{eqnarray*}

Thus, if the Lagrangian does not depend explicitly on the time variable
$t$ than the Hamiltonian function $H$ is conserved along any solution
of the Euler--Lagrange equations. 

\subsection{Hamiltonian Formalism}

The Hamiltonian Mechanics is based on a Hamiltonian function $H(t,q,p)$ over the phase-space
coordinates $(q,p)$ along with a 
Poisson bracket $\left\{ ,\right\}$ which in its canonical form~is:

\begin{equation}
\left\{ f,g\right\} =\frac{\partial f}{\partial q}\frac{\partial g}{\partial p}-\frac{\partial f}{\partial p}\frac{\partial g}{\partial q}
=-\left\{ g,f\right\} 
\label{eq:Poiasson bracket}
\end{equation}

There are two sets of Hamilton equations of motion. The first set
defines the rate of change (generalized velocity) of the coordinate
function(s) $q$:

\begin{equation}
v:=\frac{dq}{dt}=\frac{\partial H}{\partial p}=\left\{ q,H\right\} ,\label{eq:generalized velocity}
\end{equation}

the second set is equivalent to the Euler--Lagrange equations:

\begin{equation}
\frac{dp}{dt}=-\frac{\partial H}{\partial q}=\left\{ p,H\right\} ,
\end{equation}

The rate of change of any observable $f(t,q,p)$ represented as a
function over the phase space is then given by:

\begin{equation}
\frac{df}{dt}=\frac{\partial f}{\partial q}\left(\frac{dq}{dt}\right)+\frac{\partial f}{\partial p}\left(\frac{dp}{dt}\right)+\frac{\partial f}{\partial t}=\{f,H\}+\frac{\partial f}{\partial t}\label{Hamilton's evolution equation}
\end{equation}

This expression, along with the antisymmetric property of the Poisson
bracket (\ref{eq:Poiasson bracket}), makes it easy to recognize that
the Hamiltonian is conserved if $H$ does not depend explicitly on
$t$.

\subsection{Problems with the Hamiltonian Function and the Legendre Transform
for RI Systems}

The first problem that becomes clear when studying reparametrization-invariant
systems based on first-order homogeneous Lagrangians is that the 
Hamiltonian function is identically zero
\cite{Deriglazov2011,Deriglazov2016}.
The definition of a
homogeneous function of order $n$ is \cite{VGG2005,Anderson2017,Hojman1981,Rivas2001}:
\begin{equation}
v\frac{\partial f(v)}{\partial v}=nf(v)
\end{equation}
Applying this to the Hamiltonian for a homogeneous Lagrangian of order $n$ in the velocities, one has:

\begin{equation}
H=vp-L=v\frac{\partial L}{\partial v}-L=(n-1)L.
\end{equation}
Thus, the Hamiltonian will be identically zero for first-order
homogeneous Lagrangians ($n=1$) in the velocities:
\vspace{6pt}
\begin{equation}
H=vp-L=v\frac{\partial L}{\partial v}-L\Rightarrow H\equiv0.
\end{equation}

Notice that this is independent of whether one is off-shell or on-shell
(looking at solutions of the Euler--Lagrange equations of motion).
It is just saying that $L_{1}=vp$. This problem can be mitigated if one considers 
equivalent Lagrangians.
For example, $L_{(n)}=(L_{1})^{n}$ will be a homogeneous function
of order $n$. Thus, the issue ($H\equiv0)$ will not occur since 
one has $H_{(n)}=$ $(n-1)L_{(n)}$, which will be conserved if $L$
does not depend explicitly on the time coordinate $t$, implying conserved
$L$ as well, then the Euler--Lagrange equations of motion related
to various $L_{(n)}$ are equivalent to each other and correspond
to the Euler--Lagrange equations of motion for $L_{1}$:

\begin{equation}
\frac{d}{dt}\left(\frac{\partial L^{n}}{\partial v}\right)=\frac{\partial L^{n}}{\partial q}\Rightarrow\frac{d}{dt}\left(L^{(n-1)}\frac{\partial L}{\partial v}\right)=L^{(n-1)}\frac{\partial L}{\partial q},
\end{equation}
\begin{equation}
\Rightarrow\frac{d}{dt}\left(\frac{\partial L}{\partial v}\right)+(n-1)\left(\frac{d}{dt}\ln(L)\right)\left(\frac{\partial L}{\partial v}\right)=\frac{\partial L}{\partial q},
\end{equation}

Therefore, solutions for $L_{(n)}=(L_{1})^{n}$ are also solutions
for $L_{1}$ and $L_{1}=vp$ will be conserved since $L_{(n)}$ is conserved. 
However, it is not guaranteed that all solutions for 
$L=L_{1}$ would result in the conservation of $L_{1}$.

Furthermore, going between Lagrangian and Hamiltonian formulation
requires solving for $v$ in the Equation (\ref{eq:Generalized momentum})
or solving for $p$ in the Equation (\ref{eq:generalized velocity}).
When one has first-order homogeneous Lagrangian $L_{1}$ this is not
possible because there are constraints in the system of equations and the 
Hessian matrix is singular
\cite{deLeon2013,Gracia2001,Hojman1981,Teitelboim1982}. 
How to resolve the constraints have been a subject of vast research
for the purpose of achieving meaningful quantization 
\cite{1996PhRvD..53.7336L,2011PhLB..703..614P,2011PhRvD..83l5023B,
2012FoPh...42.1210G,Henneaux1992ig,Schwinger1963,Dirac1958.326D,
Dirac1958.333D,2011PhRvD..83d3503L}.
We will not follow this path here and will consider an alternative
approach to move forward towards a meaningful quantization. 
For this purpose, we now review the Canonical Quantization formalism 
\cite{Rund1966,Landau-Lifshitz1975,Pavsic2001}.

\section{From Classical to Quantum Mechanics \label{sec:From-Classical-to-QM}}

In this section, in order to briefly set the notions employed 
we follow the usual physics textbooks expositions of the quantization framework.
Here, we are not concerned with the questions of boundness of the operators 
or alternative approaches to the quantization framework; these are important technical details but are not part of the scope of the current paper. 

\subsection{Canonical Quantization}

In the standard Canonical Quantization, 
observables are replaced by operators $A\rightarrow\hat{A}$
over a Hilbert space $\mathcal{H}$, while the Poisson bracket is
replaced by a commutator. Of course, this correspondence is not a strict equality
since this would imply equality of quantum mechanics to Poissonian mechanics. Therefore, it is 
a functorial correspondence from the category of Poissonian systems to quantum systems 
where details depend on the specific systems~considered:
 
\begin{equation}
\{A,B\}\rightarrow\frac{1}{i\hbar}[\hat{A,}\hat{B}]\label{CanonicalQ}
\end{equation}
Thus, the phase-space coordinates $q$ and $p$ are viewed as operators
$\hat{q}$ and $\hat{p}$ satisfying the~relation: 

\begin{equation}
\{q,p\}=1\Rightarrow[\hat{q},\hat{p}]=i\hbar\label{qpCommutator}
\end{equation}
In particular, if the Hilbert space $\mathcal{H}$ is taken to be
the space of square integrable functions over the configuration space

\[
\mathcal{H}=\mathcal{L}_{2}[\psi(q):\hat{q}\psi(q)=q\psi(q),\intop||\psi(q)||^{2}=1]
\]
then the momentum operator $\hat{p}$ becomes: 
\vspace{6pt}
\begin{equation}
\hat{p}=-i\hbar\frac{\partial}{\partial q}\label{p as partial derivative}
\end{equation}
consistent with the notion that momentum is a generator of
translations along the coordinate $q$.

The evolution of operators is now given by the 
Heisenberg equation, which is very similar to what one had in the Hamiltonian formalism \cite{Elci2010}: 

\begin{equation}
\frac{df}{dt}=\{f,H\}+\frac{\partial f}{\partial t}\rightarrow\frac{d\hat{f}}{dt}=\frac{1}{i\hbar}[\hat{f},\hat{H}]+\frac{\partial\hat{f}}{\partial t}
\end{equation}
One shall not invoke yet the Schr\"odinger Equation since it will appear naturally in the approach presented. 
This way, one does not have to specify what $\hat{H}$ is either.

\subsection{Extending the Poisson Bracket}

Finding a
covariant formulation of the Hamiltonian Mechanics has been an 
important topic of research leading to various approaches and methods. 
Here the aim is the extension of the configuration space to include time
as a coordinate that allows us to connect to Quantum Mechanics.
Thus, one needs to define a Hamiltonian and a suitable Poisson Bracket,
and since the Hamiltonian formalism requires pairs of phase space-coordinates
$(q,p)$ then one needs to identify the corresponding partner to the time
coordinate $t$. From the theory of Special Relativity (SR), one already
knows that it should somehow be related to the energy $E$, which is
related to the $p_{0}$ component of a four-vector, and from Relativistic Quantum
Mechanics (QM) and Relativistic Quantum Field Theory (QFT) one also has: 
\begin{equation}
p_{0}\rightarrow\hat{p}_{0}=i\hbar\frac{\partial}{\partial t}
\label{Energy as translation in time}
\end{equation}

\begin{equation}
c\hat{p}_{0}\psi_{E}(t,q)=E\psi_{E}(t,q)
\end{equation}
However, measuring energy is also related to the Hamiltonian in
Classical Mechanics as well as in Quantum Mechanics. 

\begin{equation}
E=H(q,p)\rightarrow E_{\psi}=<\psi|\hat{H}|\psi>
\end{equation}
In Quantum Mechanics the Hamiltonian is often the starting point to
define a basis in the Hilbert space for the quantum system to be studied.

\begin{eqnarray*}
\mathcal{H}&=&\mathcal{L}_{2} \{\, \psi(q)=\sumint c(E)\phi_{E}(q)d\mu_E : 
H\phi_{E}(q)=E\phi_{E}(q),\intop d\mu_E||\psi(E)||^{2}=1\}
\end{eqnarray*}
However, if one starts with the first-order homogeneous Lagrangian, one has
the problem of $H\equiv0$, which is a general problem faced in
\index{other important reparametrization-invariant systems}
\textit{other important reparametrization-invariant systems}, such
as String Theory and Quantum Gravity. In 
successful modern Quantum Field Theory models, the energy $E$ is
proportional to $p_{0}$ as the generator of translations along the
laboratory time coordinate (\ref{Energy as translation in time}) while the 
Hilbert space is determined from the relevant 
Relativistic Euler--Lagrange equations, such as the 
Dirac equation or its equivalent in the relevant 
Yang-Mills theory.  So, it seems reasonable to give up on the notion that energy is related to a Hamiltonian operator 
coming from a classical system or that the basis of the Hilbert space can
be associated to a Hamiltonian operator. 
Instead, it may be better to consider that 
$H\equiv0$ should define the Hilbert space in a way similar to 
how one can derive the topology of a space by looking at the algebra of functions over this space.

We try to turn the issue $H\equiv0$ into a virtue for some suitable
expression of $\boldsymbol{H}$ that will define the Hilbert space.
\vspace{6pt}
\begin{equation}
\mathcal{H}=\{\boldsymbol{\hat{H}}\psi\equiv0\}
\end{equation}
This definition of the Hilbert space of a system clearly guarantees the principle of superposition of quantum states.
Then, the evolution of the observables for the system (process) under
a general parametrization $\lambda$ of a process studied: 
\begin{equation}
\frac{d\hat{f}}{d\lambda}=\frac{1}{i\hbar}[\hat{f},\boldsymbol{\hat{H}}]
\end{equation}
should emerge from the extension of the Hamiltonian formalism.

If one 
extends the configuration space of a classical system from ${q_{1},\ldots,q_{n}}$, to include
$t$ as $q_{0}$, to ${q_{0},q_{1},\ldots,q_{n}}$ and the phase space
by adding $p_{0}$ to the list of $(q,p)$ pairs of conjugated variables,
then one should probably extend the Hamiltonian, so that the extension
of the evolution Equation (\ref{Hamilton's evolution equation}) would
allow some meaningful interpretation. Furthermore, one would like the
extension to be useful for reparametrization-invariant
systems---so a general parametrization $\lambda$ should be used to
determine the rate of change of an observable related to a function
$f(q_{0},q_{1},\ldots,q_{n},p_{0},p_{1},\ldots,p_{n})$:

\begin{equation}
\frac{df}{dt}=\{f,H\}+\frac{\partial f}{\partial t}\rightarrow\frac{df}{d\lambda}
=\left\llbracket f,\boldsymbol{H}\right\rrbracket 
\end{equation}

\begin{eqnarray*}
\left\{ f,g\right\} &=&\frac{\partial f}{\partial q}\frac{\partial g}{\partial p}-\frac{\partial f}{\partial p}\frac{\partial g}{\partial q}\rightarrow\\
&\rightarrow& \left\llbracket f,g\right\rrbracket=\left\{ f,g\right\} +\left(\frac{\partial}{\partial x^0}\overleftrightarrow{\otimes}\frac{\partial}{\partial p_{0}}\right)\triangleright\left(f\overleftrightarrow{\otimes}g\right)
\end{eqnarray*}
This brings the attention only on the expression of the Poisson
Bracket on the $(x^0,p_{0})$ sub-space of the phase space.

One approach is to set $x^0=ct$ on a similar footing as all the other coordinates
$q_{i}$ for $i=1\ldots n$ (see for example Refs. 
\cite{Lanczos1970,Deriglazov2011,Deriglazov2016,Kuwabara198402}):

\begin{equation}
\left\llbracket x^0,p_{0}\right\rrbracket =
\left(\frac{\partial}{\partial x^0}\frac{\partial}{\partial p_{0}}-\frac{\partial}{\partial p_{0}}\frac{\partial}{\partial x^0}\right)\triangleright\left(x^0\otimes p_{0}\right)=1
\end{equation}
The rational for such a choice is to consider $q^{i}$ and $p_{j}$ as vectors living in dual spaces
and thus the natural choice $\{q^{i},p_{j}\}=\delta^i_j$ to be extended to 
$\left\llbracket q^{\mu},p_{\nu}\right\rrbracket =\delta^{\mu}_{\nu}$ which leads to
$\left\llbracket x^0,p_{0}\right\rrbracket =1$.
The problem with this choice for the particular metric signature $\{+,-,-,-\}$, 
as discussed in Section \ref{Space-Time Metric Signature},
is that there is no difference between upper and lower $0$ indexes.  
Thus, the coordinate time $t$ is too much like all the other configuration coordinates.
As if done in treatments that are using the signature $\{-,+,+,+\}$, which is easy on the index manipulations for spacial vectors.
Furthermore, there is no explicit local Lorentz invariance, and upon canonical quantization it does not give the usual expression for
$\hat{p}_{0}$ that is more like the expression for the usual space-related momenta $\hat{p}_{i}$ 
(\ref{p as partial derivative}) rather than (\ref{Energy as translation in time}). 
To resolve this issues, it seems better to consider: 
\begin{equation}
\left\{ f,g\right\} \rightarrow\left\llbracket f,g\right\rrbracket =
-\eta_{\mu\nu}\left(\frac{\partial f}{\partial q_{\mu}}\frac{\partial g}{\partial p_{\nu}}-\frac{\partial f}{\partial p_{\mu}}\frac{\partial g}{\partial q_{\nu}}\right)
\end{equation}
where $\mu=0,1,\ldots,n$ and $\eta_{\mu\nu}$ is the Lorentz-invariant
tensor with signature ${1,-1,\ldots,-1}$. The minus sign in front
of $\eta_{\mu\nu}$ is to recover the usual Poisson bracket between
$q_{i}$ and $p_{i}$ for $i=1\ldots n$ as expected in the signature $\{-,+,+,+\}$ treatment. 
This expression is explicitly Lorentz invariant, which singles out $q_{0}=ct$ as different from
the other special coordinates $q_{i}$ and it results in the correct
expression for $\hat{p}_{0}$ upon canonical quantization since now:
\vspace{6pt}
\begin{equation}
\left\llbracket x^0,p_{0}\right\rrbracket =-1
\end{equation}
Thus, the 
generalized Poisson bracket is: 

\begin{equation}
\left\llbracket f,g\right\rrbracket =\left\{ f,g\right\} -\left(\frac{\partial f}{\partial x^0}
\frac{\partial g}{\partial p_{0}}-\frac{\partial f}{\partial p_{0}}\frac{\partial g}{\partial x^0}\right)
\end{equation}

Such expression has already been derived \cite{Elci2010} based on the invariance of the
Lagrange brackets which will not be discussed here. The reader will see in the forthcoming
two examples that the meaning of the general parametrization $\lambda$
is intimately related to the choice of extended Hamiltonian $\boldsymbol{H}$.

Alternatively one can absorb the ``$\minus$'' into the $\eta_{\mu\nu}$ 
and thus adopt instead the signature $\{-,+,+,+\}$ for $\eta_{\mu\nu}$.
Then the expression $\sqrt{g(\vec{v},\vec{v})}$ in Section \ref{Space-Time Metric Signature}
will become $\sqrt{-g(\vec{v},\vec{v})}$. This is an alternative approach that seems to
arrive at very much the same general conclusions  \cite{Deriglazov2011,Deriglazov2016}.

\subsection{Implementing the Hamiltonian Constraint}

If one uses the extended Poisson bracket $\left\llbracket f,g\right\rrbracket $
one can extend also the classical Hamiltonian $H$ in a suitable way: 
\begin{equation}
H\rightarrow\boldsymbol{H}=H+?
\end{equation}
so that the Hamiltonian's evolution equations are recovered: 
\begin{equation}
\frac{df}{dx^0}=\{f,H\}+\frac{\partial f}{\partial x^0}\rightarrow\frac{df}{d\lambda}
=\left\llbracket f,\boldsymbol{H}\right\rrbracket .
\label{HamEvEq}
\end{equation}
If $\lambda$ is chosen to be the coordinate time $t$, then one has:

\begin{eqnarray*}
\frac{df}{dt}&=&\{f,H\}+\frac{\partial f}{\partial t}=\left\llbracket f,\boldsymbol{H}\right\rrbracket,\\
\left\llbracket f,\boldsymbol{H}\right\rrbracket &=&\{f,\boldsymbol{H}\}
-\left(\frac{\partial f}{\partial x^0}\frac{\partial\boldsymbol{H}}{\partial p_{0}}
-\frac{\partial f}{\partial p_{0}}\frac{\partial\boldsymbol{H}}{\partial x^0}\right).
\end{eqnarray*}
To get the usual $\frac{\partial f}{\partial t}$ that is the explicit
time derivation of an observable $f$, the 
extended Hamiltonian $\boldsymbol{H}$ should be chosen to be:

\begin{equation}
\boldsymbol{H}=H(q_{i},p_{i})-cp_{0}\label{extended Hamiltonian}
\end{equation}
Then one has:
\begin{eqnarray*}
\frac{df}{dx^0}\rightarrow\frac{df}{d\lambda}&\coloneqq&\left\llbracket f,\boldsymbol{H}\right\rrbracket 
=\{f,\boldsymbol{H}\}-\left(\frac{\partial f}{\partial x^0}\frac{\partial\boldsymbol{H}}{\partial p_{0}}
-\frac{\partial f}{\partial p_{0}}\frac{\partial\boldsymbol{H}}{\partial x^0}\right)=\\
&=&\{f,H\}-\left(\frac{\partial f}{\partial x^0}\frac{\partial(-cp_{0})}{\partial p_{0}}-\frac{\partial f}{\partial p_{0}}\frac{\partial H}{\partial x^0}\right)\\
&=&\{f,H\}+c\frac{\partial f}{\partial x^0}+\frac{\partial f}{\partial p_{0}}\left(\frac{\partial H}{\partial x^0}\right)
\end{eqnarray*}

This is the usual Hamiltonian evolution equation as long as there
is no explicit time dependence in the classical Hamiltonian $H=H(q_{i},p_{i})$,
which is often the case. Even if there is an explicit  time dependence of the classical Hamiltonian, 
the extra term will kick-in only if the observable $f$ depends explicitly on $p_0$. For example, 
if $f=p_0$ then the  extra term results in the correct expression for the rate of change of $p_0$,
The extended Poisson bracket $\left\llbracket f,g\right\rrbracket $
and the extended Hamiltonian $\boldsymbol{H}=H(q_{i},p_{i})-cp_{0}$
will recover the standard Hamiltonian formalism. Upon imposing the
requirement $\boldsymbol{H}=0$, one can determine the phase-space
of the system which will also recover the usual relationship that
the classical Hamiltonian $H(q_{i},p_{i})$ is related to the energy
of the system $H(q_{i},p_{i})=E=cp_{0}$.

An apparent drawback is that $\boldsymbol{H}=H(q_{i},p_{i})-cp_{0}$ does
not seems to be generally covariant. However, this is due to the choice
of parametrization $\lambda=x^0=ct$ that has been identified with the
laboratory time coordinate $t$. In a theory that is built upon a
reparametrization-invariant first-order homogeneous Lagrangian with
$\boldsymbol{H}\equiv0$ then $\boldsymbol{H}$ is generally covariant by~construction.

In the alternative approach developed in \cite{Deriglazov2011,Deriglazov2016}
that is using $\{q^{\mu},p_{\nu}\}=\delta^{\mu}_{\nu}$ one arrives at an extended Hamiltonian that 
has a $\plus$ sign in front of $p_{0}$, that is, $\boldsymbol{H}=H(q_{i},p_{i})+cp_{0}$.
Such expression, in general, does not convey the significance of $H\equiv 0$ because one usually expects 
the Hamiltonian $H(q_{i},p_{i})$ as well as $p_{0}$ to be related to the energy of a system 
and for non-interacting systems, all these are expected to be positive. Furthermore, 
the expression may be confused to stand for the usual addition of various energy 
components such as energy of a particle $c p_0$ and interaction energy 
with the environment $H(q_{i},p_{i})$. This could be mitigated partially if one uses the
dual of $p_0$ instead because for signature $\{-,+,...,+\}$ this results in  
$\boldsymbol{H}=H(q_{i},p_{i})-cp^{0}$.

\subsection{The Schr\"odinger Equation}

Using the above described extended Hamiltonian formalism, one can look
at the corresponding quantum picture. Most of it is already in the
correct form: 
\begin{equation}
\{A,B\}\rightarrow\left\llbracket A,B\right\rrbracket \rightarrow\frac{1}{i\hbar}[\hat{A,}\hat{B}]
\end{equation}

\begin{eqnarray}
\{q_{i},p_{i}\} =1&\rightarrow&\left\llbracket q_{i},p_{i}\right\rrbracket =1\\
&\Rightarrow& [\hat{q},\hat{p}]=i\hbar \Rightarrow\hat{p}_{i}=-i\hbar\frac{\partial}{\partial q_{i}} 
\nonumber 
\end{eqnarray}

\begin{eqnarray}
\label{p0-operator}
\{q_0,p_{0}\}=-1&\rightarrow&\left\llbracket q_0,p_{0}\right\rrbracket =-1 \\
&\Rightarrow&[\hat{q}_{0},\hat{p}_{0}]=-i\hbar \Rightarrow\hat{p}_{0}=i\hbar\frac{\partial}{\partial q_0}
\nonumber
\end{eqnarray}

\begin{equation}
\mathcal{H}=\{\boldsymbol{H}\psi\equiv0\}
\label{Hpsi=0}
\end{equation}
Looking at the expression (\ref{extended Hamiltonian}) for $\lambda=q_0=ct$, one now has: 
\begin{equation}
cp_{0}\psi=H(q_{i},p_{i})\psi
\label{p0SchEq}
\end{equation}
which is exactly 
the Schr\"odinger equation \cite{2011PhLB..703..614P}: 
\begin{equation}
i\hbar\frac{\partial\psi}{\partial t}=H(q_{i},p_{i})\psi
\label{SchEq}
\end{equation}
\newpage
One can arrive at the Schr\"odinger equation even with the alternative choice of the extended Poisson Bracket, 
and  the opposite sign in the expression of $\hat{p}_{0}$ would be handy and collaborate with $\plus$ sign in the 
alternative extended Hamiltonian ($\boldsymbol{H}=H(q_{i},p_{i})+cp_{0}$) to produce the correct Schr\"odinger equation
\cite{Deriglazov2011,Deriglazov2016}.

The Schr\"odinger Equation (\ref{SchEq}) along with its dual equation show that the norm of the state $\psi$, 
which is a solution to (\ref{SchEq}), and will be conserved as long as $H(q_{i},p_{i})$ is self-dual operator; 
that is, $H(q_{i},p_{i})$ is a Hermitian operator ($H^{\dagger}(q_{i},p_{i})=H(q_{i},p_{i})$) within the 
considered Hilbert space with the appropriate inner product. 
Thus, within the Schr\"odinger's picture of Quantum Mechanics, 
the unitary evolution of a state is a well-understood property of the corresponding picture.
The Schr\"odinger's picture of Quantum Mechanics is also equivalent to the Heisenberg's picture 
based on the Hamiltonian's evolution Equation (\ref{HamEvEq}).
In the Heisenberg's picture, the state of the system is un-mutable, while the measurement operators are 
undergoing a unitary evolution according to equation similar to Heisenberg's evolution equations \cite{landau1981quantum}.
In this section, the Schr\"odinger Equation (\ref{SchEq}) was derived upon the specification of a  particular 
mathematical expression for the extended Hamiltonian ($\boldsymbol{H}=H(q_{i},p_{i})-cp_{0}$) that is relevant
when the process parametrization $\lambda$ takes upon the meaning of coordinate time $t$ and thus 
(\ref{SchEq}) follows from (\ref{Hpsi=0})  via (\ref{p0SchEq}) and (\ref{p0-operator}). 
Under this specification, the standard Schr\"odinger's picture is recovered along with all its well-known 
unitary evolution and probability interpretation. 
The Equation (\ref{Hpsi=0}), that should be used to define the Hilbert space of the corresponding system under consideration, 
is also form-invariant upon a much larger class of unitary transformations. 
What will be the form and meaning of the unitary evolution, 
Schr\"odinger's or Heisenberg's or some other type, is not clear in general since one has to determine the meaning of the 
parameter  $\lambda$, which is intimately related to the specific mathematical expression  
that would encode the Hamiltonian constraint  (\ref{Hpsi=0}).
The connection between the specific meaning of parameter  
$\lambda$ and the mathematical form of the extended Hamiltonian $\boldsymbol{H}$
is illustrated in more detail in the following sections as well as in Appendix \ref{Appendix}.
Nevertheless, $\mathcal{H}=\{\boldsymbol{H}\psi\equiv0\}$ given by (\ref{Hpsi=0}) 
is a strong justification for the linear superposition principle that is a key property of the Quantum~Mechanics.

\section{The Meaning of \boldmath{$\lambda$} and the Role of the Hamiltonian Constraint
\label{sec:The-Meaning_of-the-process-parameter}}

In this section, we discuss the meaning of the time (evolution) parameter
$\lambda$ as related to the choice of expressing the 
Hamiltonian constraint 
of a reparametrization-invariant system based on first-order homogeneous Lagrangians in the velocities.
Up to our best knowledge, the general 
functional form of the first-order homogeneous Lagrangians
in $n$-dimensional space-time is not fully understood yet \cite{VGG2002Varna,Rund1966,Rivas2001}.
Nevertheless, since any motion of an object can be viewed either in a co-moving frame, where the object is 
practically at rest and thus moving only through time while all the other coordinates are then irrelevant, 
or one can employ curvilinear coordinates where the motion is only along one of the spatial coordinates;
that is, the motion is along its trajectory coordinate while all other spacial curvilinear coordinates are fixed.

 In this respect, the following sections still bear significant physical content and the results are valid in the 
general context as seen in  Appendix \ref{Appendix} for the case of the relativistic particle;
however, we prefer the main exposition below to be without the unnecessary clutter of multi-dimensional notations.

\subsection{The Picture from Lagrangian Mechanics' Point of View}

For the simplest possible case of 
only one space-time coordinate $q$, 
one has an explicit unique form for the Lagrangian based on the Euler's
equation for homogeneous functions of the first-order in the velocity:
\begin{equation}
v\frac{\partial L(q,v)}{\partial v}=L(q,v)\Rightarrow L(q,v)=\phi(q)v
\end{equation}
The action $\mathcal{A}$ will take a very simple form: 
\begin{equation}
\mathcal{A}=\int L(q,v)d\lambda=\int\phi(q)vd\lambda=\int\phi(q)dq
\label{Lagrangian example 1}
\end{equation}
The Euler--Lagrange equations are now: 
\begin{equation}
\frac{dp}{d\lambda}=\frac{\partial L}{\partial q},\:p\coloneqq\frac{\partial L}{\partial v}=\phi(q)
\label{ELE4q-space}
\end{equation}
The Hamiltonian function is then: 
\begin{equation}
H=pv-L\equiv0
\end{equation}

At this point, there are two choices for the meaning of the coordinate $q$. 
It could be a spatial coordinate or a time coordinate. 
For the present exposition, the time-like coordinate is of special interest, 
but as a warmup and for comparison we first discuss  the space-like~case.

\subsubsection{The Proper Length Parametrization and the Onset of Quantum Length Scale}

If one chooses to associate $q$ with a position in space then $v$ can
be the coordinate velocity if $\lambda=t$. In general, one has $v=dq/d\lambda$
and looking for the $q(\lambda)$ that minimizes $\mathcal{A}$ tells
us a trajectory that has a unique value associated to it---the minimum
$\Delta$ for the action $\mathcal{A}$ where $q(\lambda)$ satisfies
the Euler--Lagrange equations. However, the Equation (\ref{ELE4q-space}) 
are now trivially satisfied for any $q(\lambda)$ since $L(q,v)=\phi(q)v$, and $p=\phi(q)$:
\begin{equation}
\frac{d\phi(q)}{d\lambda}=\frac{\partial\phi}{\partial q}v
\Leftrightarrow
\frac{\partial\phi}{\partial q}\frac{dq}{d\lambda}=\frac{\partial\phi}{\partial q}v
\end{equation}
This is very similar to the multi-dimensional case discussed in Reference  \cite{Bouvier197804}.
The Hamiltonian function is not telling us anything new either,
it is just bringing us back to the original expression for $L=pv$ with $p=\phi(q)$. 

However, based on conservation of the equivalent Lagrangians 
$L_{(n)}=\left(L_{1}\right)^{n}$ one can impose
$\frac{dL_{1}}{d\lambda}=0$ to ensure conservation of 
$L_{1}$ that gives us an additional equation: 
\begin{eqnarray}
\frac{dL_{1}}{d\lambda}=0
\Rightarrow\frac{d\phi(q)}{d\lambda}v+\phi(q)\frac{dv}{d\lambda}=0\\
\Rightarrow\frac{dv}{d\lambda}=-v^{2}\frac{\partial\ln\phi(q)}{\partial q}
\label{eq:v and phy for L equal to a const}
\end{eqnarray}

The equation for the rate of change of the velocity $v$ is gauge
invariant under the general change of parametrization 
$\lambda\rightarrow\xi$ given by a general function $\lambda(\xi)$: 
\begin{equation}
\frac{d(v^{-1})}{d\lambda}=\frac{\partial\ln\phi(q)}{\partial q}
\Leftrightarrow
\frac{d(w^{-1})}{d\xi}=\frac{\partial\ln\psi(q)}{\partial q}\label{eq:gauge invariant equation for v and phy}
\end{equation}
where $v=\frac{dq}{d\lambda}$, $w=\frac{dq}{d\xi}$, and 
$\phi(q)=\psi(q)\lambda'(q),$
with $\lambda'(q)=\left.\frac{d\lambda}{d\xi}\right|_{q}$.

The gauge invariance can be seen either from the expression for the 
Lagrangian $L$ within the action $\mathcal{A}$ (\ref{Lagrangian example 1}) 
or from the following direct mathematical considerations.

Consider first how the left-hand side transforms: 
\begin{eqnarray*}
\frac{d(v^{-1})}{d\lambda}&=&\frac{d}{d\lambda}\left(\frac{d\lambda}{dq}\right)\rightarrow\\
&=&\frac{d}{d\lambda(\xi)}\left(\frac{d\lambda(\xi)}{dq}\right)=\frac{d}{\lambda'd\xi}\left(\frac{\lambda'd\xi}{dq}\right)=\\
&=&\frac{d}{d\xi}\left(\frac{d\xi}{dq}\right)+\frac{\lambda''}{\lambda'}\left(\frac{d\xi}{dq}\right)\\
&=&\frac{d}{d\xi}\left(w^{-1}\right)+\frac{d\ln(\lambda')}{d\xi}\left(\frac{d\xi}{dq}\right)\\
&=&\frac{d}{d\xi}\left(w^{-1}\right)+\frac{d\ln(\frac{d\lambda}{d\xi})}{dq}
\end{eqnarray*}

\begin{equation}
\Rightarrow\frac{d(v^{-1})}{d\lambda}\rightarrow\frac{d}{d\xi}\left(w^{-1}\right)+\frac{d\ln(\lambda')}{dq}
\end{equation}
the right-hand side transforms as follows: 
\vspace{6pt}
\begin{equation}
\frac{\partial\ln\phi(q)}{\partial q}
\rightarrow\frac{\partial\ln(\psi(q)\lambda')}{\partial q}
=\frac{\partial\ln(\psi(q))}{\partial q}+\frac{\partial\ln(\lambda')}{\partial q}
\end{equation}
Thus, the second ($\ln(\lambda')$) terms would cancel out.

By looking at $\phi(q)$ in the Equation (\ref{eq:v and phy for L equal to a const})
one can see that if $\phi(q)$, which is the momentum $p$, is constant
($p=\phi(q)=\phi_{0})$ during the process then $v$ does not change
and is conserved along the path $q(\lambda)=v\lambda+q(0)$ and the
constant value of the Lagrangian is $\phi_{0}v$ $(L=\phi_{0}v)$.
If one assumes $\lambda=v^{-1}q$ with $v$ an arbitrary non-zero constant then the
Equation (\ref{eq:v and phy for L equal to a const}) demands $\phi(q)$,
the momentum $p$, to be constant $p=\phi(q)=\phi_{0}$: 
\begin{equation}
0=\frac{d(v^{-1})}{d\lambda}=\frac{\partial\ln\phi(q)}{\partial q}
\end{equation}

In this case a new parametrization can be chosen---the proper-length
$\mathit{l}$ so that\linebreak ($dl=\phi_{0}dq=\phi_{0}vd\lambda$) and the
value of the Lagrangian becomes one
 ($L=\phi_{0}v\rightarrow L=1$): 
\begin{eqnarray*}
\mathcal{A}&=&\int L(q,v)d\lambda=\int\phi(q)vd\lambda=\int\phi(q)dq \\
&\Rightarrow&\Delta\mathit{l=}\int\phi_{0}dq=\phi_{0}\Delta q=\phi_{0}v\Delta\lambda
\end{eqnarray*}
The proper-length 
 $\mathit{l}$ can be introduced even in the general
case of $\phi(q)$: 
\begin{eqnarray*}
dl&=&\phi(q)dq=\phi(q)vd\lambda ,\\
\Rightarrow\mathcal{A}&=&\int L(q,v)d\lambda=\intop dl=\Delta l
\end{eqnarray*}

If one chooses the proper length $dl=\phi(q)dq$ as parametrization,
then the Lagrangian is explicitly a constant ($L=1$) and the ``velocity''
is $w=dq/dl=\frac{1}{\phi(q)}$, but from the form invariant expression
of the action integral, $\psi(q)wdl=\phi(q)vd\lambda=\phi(q)dq=dl$
one also has $L=pw$, which requires $p=\psi(q)=\phi(q)$ so that $L=1$.
In this case the Euler--Lagrange~equations: 
\begin{equation}
\frac{dp}{dl}=\frac{\partial L}{\partial q},\:p\coloneqq\frac{\partial L}{\partial w}=\psi(q)=\phi(q)
\end{equation}
are trivially satisfied as well: 
\begin{equation}
\frac{d\phi(q)}{dl}=\frac{\partial L}{\partial q}\rightarrow\frac{d\phi(q)}{dq}\frac{dq}{dl}=w\frac{\partial\phi(q)}{\partial q}\rightarrow\frac{1}{\phi(q)}\frac{\partial\phi(q)}{\partial q}
\end{equation}

If one looks for any other parametrizations, which correspond to constant
$L$ and thus satisfy (\ref{eq:v and phy for L equal to a const})
or the equivalent Equation (\ref{eq:gauge invariant equation for v and phy}),
one can conclude that there is a family of parametrizations up to
a constant factor $\lambda_{0}$ related to $\lambda_{0}d\lambda=\phi(q)dq$. 
For this purpose, consider (\ref{eq:gauge invariant equation for v and phy})
such that $\lambda=\lambda(q)$ since there is no other variable for
$\lambda$ to depend on. Then Equation (\ref{eq:gauge invariant equation for v and phy})
gives us: 
\begin{eqnarray*}
\frac{\partial\ln\phi(q)}{\partial q}&=&\frac{d(v^{-1})}{d\lambda}=\frac{d}{d\lambda}\left(\frac{d\lambda}{dq}\right)\\
&=&\left(\frac{dq}{d\lambda}\right)\frac{d}{dq}\left(\frac{d\lambda}{dq}\right)=\frac{d\ln\lambda'(q)}{dq}
\end{eqnarray*}

\[
\Rightarrow\frac{\partial\ln\phi(q)}{\partial q}=\frac{d\ln\lambda'(q)}{dq}
\]
with the general solution $\lambda'(q)=\phi(q)/\lambda_{0}\Rightarrow\lambda_{0}d\lambda=\phi(q)dq$.

The above considerations show that there is always a choice of parametrization
that makes the Lagrangian constant. In particular, in the proper-length
parametrization when using $dl=\phi(q)dq$ one can make $L=1$. 
Of course, there is also the question whether $\phi(q)$ is well behaved
in order to establish good equivalence between the $l$ value and
the $q$ coordinate of the trajectory of the process with $p=\phi(q)$.

Since $p=\phi(q)$ should be related to the momentum of the system,
then a process can be considered 
classical with conserved linear momentum if a value can be associated
to the process and it is independent of the observational length scale
$\Delta q$ via the following expression: 
\begin{equation}
p(\Delta q)\coloneqq\left\langle \phi\right\rangle _{\Delta q}
=\frac{1}{\Delta q}\intop_{0}^{\Delta q}\phi(q)dq
=\frac{1}{\Delta q}\intop_{0}^{\Delta l}dl=\frac{\Delta l}{\Delta q}
\label{eq:Conservation of momentum-Statistically}
\end{equation}
The simplest examples for such averaging with observational length scale are the density of a material,
the large scale structure of the Universe, or the average speed of a city bus, or train, etc.

If $p(\Delta q)=\phi(q)=\rm const$, then the coordinate $q$ is proportional to the proper length $l$.
In particular, in the center of mass of a system one can expect $p=0$ which will
mean that there is no change in $l$ for the process. However, this
may also happen due to quantum effects when  quantum fluctuations are
canceling out beyond a large enough scale $\Delta$. Furthermore, if one studies
natural processes at shorter and shorter scales than one may encounter
systems where the proper-length is poorly defined due to fluctuations
of $\phi(q)$ and the above formula is not applicable because of limitations
at very small scales. 
The observation of such a length scale $\delta$ can signal 
the  onset of quantum phenomenon. 

\subsubsection{The Proper Time Parametrization and the Onset of Quantum Time Scale}

Now, let us consider the possibility that the $q$ coordinate is time-like.
In what follows, $q$ will be set to be the laboratory time coordinate
$t$ and the rate of its change $dt/d\lambda$ will be denoted with
$u$ instead of $v_{0}$ but we will not use $E$ nor $p_{0}$ for
$p$ which in this case carry the correct meaning of $p$: 
\begin{equation}
L_{1}(t,u)=\phi(t)u\Rightarrow u\coloneqq\frac{dt}{d\lambda},\:p\coloneqq\frac{\partial L_{1}}{\partial u}=\phi(t),
\end{equation}
\begin{equation}
\frac{dp}{d\lambda}=\frac{\partial L_{1}}{\partial t}\Rightarrow\frac{d\phi(t)}{d\lambda}=u\frac{\partial\phi(t)}{\partial t}
\end{equation}

The corresponding Hamiltonian function is then: 
\begin{equation}
H=pu-L\equiv0
\end{equation}
but one cannot say anything about the rate with which $u$ is changing.
The action $\mathcal{A}$ will take the value $\Delta$ for the overall
observed motion: 
\begin{equation}
\mathcal{A}=\int L(t,u)d\lambda=\int u\phi(t)d\lambda=\int\phi(t)dt=\Delta
\end{equation}

Since the model is reparametrization invariant, one can define a quantity
that different observers can deduce from observations and compare---this is the proper-time parametrization $\tau$: 
\begin{equation}
d\tau=\phi(t)dt\label{tau-phi-t}
\end{equation}

In this parametrization the action $\mathcal{A}$ will take simpler
form: 
\begin{equation}
\mathcal{A}=\int L(t,u)d\lambda=\int u\phi(t)d\lambda=\int\phi(t)dt=\int d\tau=\Delta\tau
\end{equation}
and different observers will be able to compare different phases of
the process and deduce overall scale factor that will allow identical
results.

Furthermore, for the equivalent Lagrangians $L_{(n)}=\left(L_{1}\right)^{n}$
there is an explicit time dependence. Thus, the corresponding Hamiltonian
functions will not be integrals of the motion. For example, $H_{(2)}=(L_1)^{2}=\phi(t)^2u^2$.
However, the proper-time parametrization will make $L_{1}=1$ or by
requiring $\frac{dL_{1}}{d\lambda}=0$ one will arrive again at: 
\begin{equation}
\frac{dL_{1}}{d\lambda}=0\Rightarrow\frac{du}{d\lambda}=-u^{2}\frac{d\ln\phi(t)}{dt}
\end{equation}
which has the general solutions $\lambda_{0}d\lambda=\phi(t)dt$ as discussed
in the previous section.

However, again if $\phi(t)=\phi_{0}$ is a constant then $u=\zeta$
is a constant too; therefore, the rate of change of $t$ and $\lambda$
are proportional. This means that one can choose the unit of the process
time $\lambda$ to be the same as the coordinate time $t$ which makes
$u=1$. Therefore, the action integral will give us: 
\begin{equation}
\mathcal{A}=\int L(t,u)d\lambda=\int\phi(t)dt=\Delta\tau\rightarrow\phi_{0}\int dt=\phi_{0}\Delta t=\phi_{0}\zeta\Delta\lambda
\end{equation}

Alternatively, if one starts with $\lambda=t$, then one has $u=1$,
$L=\phi(t)$, and $p$ should be assumed to be $\phi(t)$ since that
was the case for all other choices of parameterization. Then one can
consider the proper time $\tau$ as a new choice of parametrization
to study the system. In the proper-time parametrization $L=1$, which
is explicitly a constant. It seems that for massive particles/systems
one can always expect $L$ to be non-zero and thus in the proper-time
parametrization to be set to 1. Since the corresponding momentum $p=p_{0}=\phi(t)$ should be
related to the energy of the system, then a process can be considered
classical with conserved energy if a quantity (energy) can be associated
to the process and it is independent of the observational time interval
$\Delta t$ via: 
\begin{equation}
E=E(\Delta t)\coloneqq\left\langle \phi\right\rangle _{\Delta t}
=\frac{1}{\Delta t}\intop_{0}^{\Delta t}\phi(t)dt=\frac{1}{\Delta t}\intop_{0}^{\Delta\tau}d\tau
=\frac{\Delta\tau}{\Delta t}\label{eq:Conservation of Energy-Statistically}
\end{equation}

However, if one studies natural processes at shorter and shorter time
scales then one may encounter systems where the proper-time is poorly
defined due to fluctuations of $\phi(t)$ and the above formula is
not applicable because of fluctuations at very small time scales.
The observation of such time scale $\delta$
can signal the onset of quantum phenomenon.

\subsection{The Picture from Hamiltonian Mechanics Point of View}

Consider now the same system but from the Hamiltonian point of view
using the extended Poisson bracket. The main relationships in the
Lagrangian formulation based on the Lagrangian $L_{1}(t,u)=\phi(t)u$
are $u\coloneqq\frac{dt}{d\lambda}$, $p\coloneqq\frac{\partial L_{1}}{\partial u}=\phi(t)$,
and $H=pu-L\equiv0$. If one considers the choice of parametrization
$\lambda$ to be the laboratory time coordinate $t$ then one has
$u=1$, $L_{1}(t,u)=\phi(t)$, and $H=p-L\equiv0$, which is consistent
with the general expression $p\coloneqq\frac{\partial L_{1}}{\partial u}=\phi(t)$
that holds for general choice of parameterizations. Thus the general
constraint would be to make sure that (in what follows we will use $c=1$ units):

\begin{equation}
p_{0}=\phi(t)\label{p0 and phy constraint}
\end{equation}
Here we use explicitly the sub-index zero to emphasize that this is
to be related to the energy momentum of a system. 

\subsubsection{Hamiltonian Constraint for $\lambda$ in Coordinate-Time Role ($\lambda=t$)}

The above expression immediately suggests an extended Hamiltonian
in the spirit of $H\rightarrow\boldsymbol{H}=H-p_{0}$ that will have
the form: 
\begin{equation}
\boldsymbol{H}=\phi(t)-p_{0}
\end{equation}
Now an interesting question is: How the phase-space coordinates evolve
and what is the meaning of $\lambda$ for such choice of extended Hamiltonian? 
To answer this question one looks at the evolution equation
for the function $t$ and for $p_{0}$: 
\begin{equation}
\frac{dt}{d\lambda}=\left\llbracket t,\boldsymbol{H}\right\rrbracket \Rightarrow\frac{dt}{d\lambda}=\left\llbracket t,(-p_{0})\right\rrbracket =1
\end{equation}
Thus this immediately tells us that the choice of $\lambda$ is actually
the laboratory time coordinate. Now one has to confirm the consistency
by looking at the evolution of $p_{0}$:

\begin{equation}
\frac{dp_{0}}{d\lambda}=\left\llbracket p_{0},\boldsymbol{H}\right\rrbracket \Rightarrow\frac{dp_{0}}{d\lambda}=\left\llbracket p_{0},\phi(t)\right\rrbracket =\frac{\partial\phi(t)}{\partial t}
\end{equation}
Thus, the choice of $\boldsymbol{H}=\phi(t)-p_{0}\equiv0$ corresponds
to $\lambda=t$ indeed.

\subsubsection{Hamiltonian Constraint for $\lambda$ in the Proper-Time Role ($\lambda=\tau$)}

The constraint in Equation (\ref{p0 and phy constraint}) has
many possible realizations. Another possibility~is: 
\begin{equation}
\boldsymbol{H}=1-\frac{p_{0}}{\phi(t)}\label{H4tau}
\end{equation}
What is the meaning of $\lambda$ for this form of $\boldsymbol{H}$? 
Again one looks at the evolution of $t$ and $p_{0}$: 
\begin{equation}
\frac{dt}{d\lambda}=\left\llbracket t,\boldsymbol{H}\right\rrbracket \Rightarrow\frac{dt}{d\lambda}=\frac{1}{\phi(t)}\left\llbracket t,(-p_{0})\right\rrbracket =\frac{1}{\phi(t)}
\end{equation}
Thus, this is the proper time parametrization choice since $d\lambda=\phi(t)dt=d\tau$.
Again one checks the consistency by looking at $p_{0}:$ 
\begin{eqnarray*}
\frac{dp_{0}}{d\lambda}=\left\llbracket p_{0},\boldsymbol{H}\right\rrbracket \Rightarrow\frac{dp_{0}}{d\lambda}
=-p_{0}\left\llbracket p_{0},\frac{1}{\phi(t)}\right\rrbracket \\
\Rightarrow-p_{0}\frac{\partial\left(\phi(t)\right)^{-1}}{\partial t}=-p_{0}\left(\frac{-1}{\phi(t)^{2}}\right)\frac{\partial\phi(t)}{\partial t}
\end{eqnarray*}
Since one has to keep $p_{0}=\phi(t)$ this finally gives the same
expression as in the laboratory coordinate time because $d\lambda=\phi(t)dt=d\tau$:
\begin{equation}
\frac{dp_{0}}{d\tau}=\frac{1}{\phi(t)}\frac{\partial\phi(t)}{\partial t}\Rightarrow\frac{1}{\phi(t)}\frac{dp_{0}}{dt}=\frac{1}{\phi(t)}\frac{\partial\phi(t)}{\partial t}
\end{equation}

\subsubsection{The Quantum Mechanics Picture and the Positivity of the Energy \label{subsec: Positivity of the Energy}}

If one applies the Canonical Quantization formalism to the extended Hamiltonian
framework with the $\boldsymbol{H}$ for the time coordinate parametrization
$\lambda=t$, one obtains the standard Schr\"odinger equation: 
\begin{equation}
\boldsymbol{H}=\phi(t)-p_{0}\rightarrow\boldsymbol{\hat{H}}\psi(t)
=0\Rightarrow i\hbar\frac{\partial\psi}{\partial t}=\phi(t)\psi
\end{equation}
where the wave function solutions $\psi(t)$ are given by: 
\begin{equation}
\psi(t)=\frac{1}{\mathcal{N}}\exp\left[-\frac{i}{\hbar}\left(\intop_{0}^{ t<\delta}\phi(t)dt
+p_{0}\intop_{\delta}^{ t\gg\delta}dt\right)\right]
\label{psi4H4tau}
\end{equation}

In the above expression we have tried to emphasize the two possible regimes for $\phi(t)$. 
In the second part the integral $\intop_{\delta}^{ t\gg\delta}\phi(t)dt$ is replaced with
its corresponding expression containing the value $p_0$ defined by (\ref{eq:Conservation of Energy-Statistically}) 
for observational window $\Delta{t}\gg\delta$ when one expects~(\ref{p0 and phy constraint}) to be valid.
When this term is dominant, which is when the coordinate-time interval of the process $\Delta t\gg\delta$ 
such that the energy $p_{0}=E$ is conserved, as discussed earlier in the
Lagrangian formulation of this system (\ref{eq:Conservation of Energy-Statistically}),
then this is the familiar plane wave with normalization factor $\mathcal{N}=1$.
However, for fluctuation of $\phi(t)$ at short time scale $\Delta t\apprle\delta$, 
which is the first integral in the above expression, that does not show energy conservation for the process, 
then the wave function is related to the integral of $\phi(t)$ 
and the normalization $\mathcal{N}$ may now depend on the size of $\delta$ and 
the structure of the relevant Hilbert space and the second term may not be present when $\delta\rightarrow\infty$. 
The value of $\delta$ depends on where there is a scale beyond which $p_0$ is conserved; 
therefore,  $\delta$ if it exists, it is a system/process specific.
It is a standard procedure to choose $\mathcal{N}$ to be a positive real number that guarantees 
the state wave function to be normalized to 1 for the chosen inner product \cite{landau1981quantum}.
Generally, the 
inner product in the space of solutions that turns it into a Hilbert space could be tricky
and may need appropriate extension of the notion of model space 
beyond the standard Hilbert space framework \cite{2019JPhCS1194a2007A}.
Regardless of the particular choice of the inner product,  one can always determine 
$\mathcal{N}>0$ based on the fact that the inner product results in an appropriate norm 
with measure that is positive $\left\Vert \psi\right\Vert ^{2}=\left\langle \psi|\psi\right\rangle>0$
due to the general properties of the inner product.

For the current purpose, however, a running average may be useful. In particular,
for the plane waves one can use a standard mathematical scalar product where $\Delta$
would be a sufficiently long observational window for the process
such that energy is conserved and thus slight variations in the window
time duration $\Delta$ are producing consistent results. 
Having consistent results is an important assumption here
and in this sense is related to the details of the chosen inner-product: 

\begin{equation}
\left\langle \Psi|\Phi\right\rangle _{\Delta}
=\frac{1}{\Delta}\intop_{t_0}^{t_0+\Delta}\Psi^{*}\Phi dt 
\label{t-inner-product}
\end{equation}

The interpretation of $\Psi^{*}\Psi$ as probability density over the configuration space 
(extended space-time configuration space) is still valid. For example, by including a spatial volume $V$
integration in (\ref{t-inner-product}) one arrives at a well-normalized state in the case of 
``standard quantum-mechanical'' Newtonian-time description of a system where the states of the system 
are already specially normalized to $1$ such that the volume integral satisfies $\intop_{V}{\Psi^{*}\Phi d^3{x}} =1$. 
In this case, one extends trivially the integration to be over $V\times \left[t_0,t_0+\Delta\right] $ 
and one can see the need of the factor $\frac{1}{\Delta}$ to complete the normalization of the state to $100\%$
probability. That is, doing a measurement within and time window  $\left[t_0,t_0+\Delta\right]$ results in observing 
the particle somewhere in the space $V$ at some moment between $\left[t_0,t_0+\Delta\right]$ during the measurement.
Even more, the above expression is still reasonable even in the case of plane-waves where
the usual quantum mechanics has issues in coming up with a well-normalizable state formulation.
In the case of the plane-waves one may have to consider also the extend of the spacial measurement 
as part of the $\Delta$ factor.
Here the  time duration $\Delta$ is the (spacetime) window of an experimental measurement process that plays 
the role of regularization procedure.  In some sense, it is related to the resolution of the measurement 
since any two plane waves with periods $T_m$ and $T_n$ such that $(T_m T_n)/|T_m-T_n|<<\Delta$
will be practically orthogonal. For example, consider a Fourier series based on periods $T_n=\varepsilon/n$
with $\varepsilon<<\Delta$ then one has that all members of the series form orthonormal basis 
since $\varepsilon/|m-n|<<\Delta$ as long as one considers $\varepsilon/\Delta\approx 0$.
Finally, the specific value $t_0$ is more of a place holder for when the measurements where made, 
but the outcome of identical experiments should not really depend on it as long as all the external conditions
have been also  $t_0$---independent.

In the case of the proper-time parametrization $\lambda=\tau$ one
is facing the question of ordering of the operators that can be resolved
by the requirement of Hermiticity of the extended Hamiltonian with
respect to the usual QM rules: 
\begin{equation}
\boldsymbol{H}=1-\frac{1}{2}\left(\frac{1}{\phi(t)}p_{0}+p_{0}\frac{1}{\phi(t)}\right)
\end{equation}
The corresponding Schr\"odinger like equation now will have an additional term: 
\begin{equation}
\psi(t)-\frac{1}{2}\left(\frac{1}{\phi(t)}\hat{p}_{0}\psi(t)+\hat{p}_{0}\frac{\psi(t)}{\phi(t)}\right)=0
\end{equation}
\begin{equation}
\psi(t)-\frac{i\hbar}{2}\frac{1}{\phi(t)}\left(\frac{\partial\psi}{\partial t}+\phi(t)\frac{\partial}{\partial t}\left(\frac{\psi(t)}{\phi(t)}\right)\right)=0
\end{equation}
\begin{equation}
i\hbar\frac{\partial\psi}{\partial t}=\left[\phi(t)+\frac{i\hbar}{2}\left(\frac{\partial\ln\phi(t)}{\partial t}\right)\right]\psi(t) \label{SchInRestFrame}
\end{equation}

Therefore, the amplitude of the original plane wave will be modulated
now by an additional factor $\varrho(t)$ satisfying: 
\begin{equation}
\frac{\partial\varrho(t)}{\partial t}=\frac{1}{2}\left(\frac{\partial\ln\phi(t)}{\partial t}\right)\varrho(t)
\end{equation}
This factor will not disappear for $\Delta t\gg\delta$ when the energy
$p_{0}=E$ is conserved. It will have the form $\varrho(t)=\sqrt{\phi(t)}$
and now the wave function will be: 
\begin{equation}
\psi(t)=\frac{1}{\mathcal{N}}\sqrt{\phi(t)}
\exp\left[-\frac{i}{\hbar}\left(\intop_{0}^{t<\delta}\phi(t)dt+p_{0}\intop_{\delta}^{t\gg\delta}dt\right)\right]
\label{wave function psi(t)}
\end{equation}
The expression (\ref{wave function psi(t)}) shows that the complex conjugated wave function $\psi(t)^{*}$ 
should, therefore, be viewed as the wave function for the time reversal process of the original process.
Notice that $\phi(t)$ is expected to be positive in order to be physical, which guarantees proper causal 
relationship as indicated by the relationship between the proper-time and the coordinate-time (\ref{tau-phi-t})
and discussed in Section \ref{positive-mass}.
Furthermore, since (\ref{SchInRestFrame}) is a linear equation for the wave function $\psi(t)$, the solution is determined 
up to an overall scalar factor.  It is a standard procedure to choose $\mathcal{N}$ to be a positive real number 
that guarantees the wave function to be normalized to 1 for the chosen inner product, 
in the example case considered, the inner product is given by (\ref{t-inner-product}).

Furthermore, for processes when energy conservation is observed
the modulating factor modifies the wave function normalization to 
$\mathcal{N}=\sqrt{p_{0}}$.
The result is very interesting since 
\textbf{the positivity of the norm now requires positivity of the energy} 
$E=p_{0}>0$ since $\phi(t)\rightarrow p_{0}$.
In the rest frame this should correspond to the rest mass of the particle.

\begin{equation}
\left\Vert \psi\right\Vert ^{2}=\left\langle \psi|\psi\right\rangle _{\Delta}
=\frac{1}{\mathcal{N}^{2}\Delta}\intop_{0}^{\Delta\gg\delta}\phi(t)dt
\underset{\Delta\rightarrow\infty}{\longrightarrow}\frac{p_{0}}{\mathcal{N}{}^{2}}>0.
\label{positivity_of_p0}
\end{equation}
Since the normalization factor $\mathcal{N}$  is somewhat arbitrary one may choose $N=1$ and keep track 
of the overall norm of the state $\left\Vert \psi\right\Vert^{2}$, in this case one has $\left\Vert \psi\right\Vert^{2}=p_0$.
Thus, the the positivity of the norm $\left\Vert \psi\right\Vert^{2}>0$ implies positivity of the energy $E=p_{0}>0$. 
Notice that somewhat similar result has been obtained in Reference  \cite{Deriglazov2011}, where 
the quantum-mechanical probability density has been related to the energy density of the wave-function potential.

By applying the operator $\hat{p}_0$ on the wave function (\ref{wave function psi(t)}) 
and considering the limit $\phi(t)\rightarrow p_{0}>0$, one has:
\begin{equation}
\hat{p}_0\psi(t)=i\hbar\frac{\partial}{\partial t}\psi(t)=
\left(\frac{i\hbar}{2}\frac{\partial \ln\phi(t)}{\partial t}+\phi(t)\right)\psi(t)\rightarrow p_{0}\psi(t)
\end{equation}
and by doing the same on the complex conjugated function one has:
\begin{equation}
\hat{p}_0\psi(t)^*=i\hbar\frac{\partial}{\partial t}\psi(t)^*=
\left(\frac{i\hbar}{2}\frac{\partial \ln\phi(t)}{\partial t}-\phi(t)\right)\psi(t)^*\rightarrow -p_{0}\psi(t)^*
\end{equation}
Thus, the complex conjugate wave function corresponds to well-normalized but negative energy state 
that could also be viewed as a time-reversal state with positive energy since $p_0>0$ in both cases:
\begin{eqnarray}
\psi(t)^*&=&\frac{1}{\mathcal{N}}\sqrt{\phi(t)}
\exp\left[+\frac{i}{\hbar}\left(\intop_{0}^{t<\delta}\phi(t)dt+p_{0}\intop_{\delta}^{t\gg\delta}dt\right)\right]= \label{first expression} \\
&=&\frac{1}{\mathcal{N}}\sqrt{\phi(t)}
\exp\left[-\frac{i}{\hbar}\left(p_{0}\intop^{\delta}_{\infty}dt+\intop^{0}_{\delta}\phi(t)dt\right)\right]
\label{wave function psi(t)^*}
\end{eqnarray}
The first expression (\ref{first expression}) above  is the complex conjugated function $\psi(t)$ in (\ref{wave function psi(t)}), 
while the second expression is the time reversal of $\psi(t)$ in (\ref{wave function psi(t)}) as encoded by the order of integration. 

\textls[-18]{In  Appendix \ref{Appendix}, the multi-dimensional case of the relativistic particle has been considered.
The results corresponding to the above discussion are given by Equations (\ref{rpspnorm1}) and (\ref{rpspnorm_gamma}).} 
As seen from the discussion in  Appendix \ref{Appendix}, there is a preferred choice of an inner product that results in a normalized state 
when the size of measurement window $\Delta_\tau$ is chosen to correspond to the proper-time interval for the measurement. 
In general, however, an observer could define the inner product based on the measurement window $\Delta_t$ 
corresponding to the coordinate-time $t$ in the lab, in this case, the norm of the state is related to the 
relativistic factor $\gamma=1/\sqrt{1-v^2/c^2}$. 
The relativistic factor $\gamma$ is usually considered to be positive, and it should be positive since according to the result in  
(\ref{rpspnorm_gamma}) it is related to the norm of the state. The relativistic factor can be expressed also in terms of 
the energy-momentum components and in the particular case of conserved energy-momentum, the positivity of the energy 
follows since $p^{0}=\gamma\sqrt{-g_{\mu\nu}p^{\mu}p^{\nu}}$. 

\subsubsection{The Rate of Change along a Coordinate and Normalizability of the Wave Function}

The extended canonical Poisson brackets were chosen to result in the
Lorentz-invariant bracket that gives us the usual form for the momentum
operators in quantum mechanics along with a reasonable new expression
for the evolution equation that involves the extended Hamiltonian
$\boldsymbol{H}$. In analogy to the evolution equation, one can ask
what is the meaning of $\lambda$ if one decides to choose $\boldsymbol{H}$
to be any of the linear momentum generators? If one does so, then one
sees that $\boldsymbol{H}=p_{1}$ in the evolution equation corresponds
to change along the $q_{1}$ coordinate since then $d\lambda=dq_{1}$
(here 1 indicates any of the spatial coordinates): 
\begin{equation}
\frac{df}{d\lambda}=\left\llbracket f,\boldsymbol{H}\right\rrbracket \rightarrow\frac{df}{d\lambda}
=\left\llbracket f,p_{1}\right\rrbracket \Rightarrow\frac{dq_{1}}{d\lambda}
=\left\llbracket q_{1},p_{1}\right\rrbracket =1
\end{equation}
For $f=p_{1}$ this will give us conservation of $p_{1}$: 
\begin{equation}
\frac{df}{d\lambda}=\left\llbracket f,\boldsymbol{H}\right\rrbracket \rightarrow\frac{dp_{1}}{d\lambda}
=\left\llbracket p_{1},p_{1}\right\rrbracket \Rightarrow\frac{dp_{1}}{d\lambda}=0
\end{equation}
If one tries to construct the phase space of this system using $\boldsymbol{H}=0$
one will get only $p_{1}=0$ which is expected to correspond to non-moving
particle along $q_{1}$. To get to the more accurate expression one must
take advantage of the fact that $\boldsymbol{H}$ relevant for the
evolution equation is determined up to a constant. That is, $\boldsymbol{H}$
and $\boldsymbol{H+const}$ will give us the same evolution equations.
Thus, in the above example it is more relevant to consider $\boldsymbol{H}=p_{1}-p_{1}(0)$.
The meaning of the constant becomes clear from the condition $\boldsymbol{H}=0$
on the states to be considered; then the $constant$ is the value
of the conserved momentum $p_{1}(0)$. This is in agreement with the
discussion on the Lagrangian $L=\phi(q)(dq/d\lambda)$ when $q$ was
a spatial coordinate and demanded constant value of $\phi(q)=\phi_{0}=p_{1}(0)$
by the choice of parametrization $q=\lambda+q(0)$. Furthermore,
since $\boldsymbol{H}=p_{1}-\phi_{0}$ can also be viewed as equivalent
to $\boldsymbol{H}=v(p_{1}-\phi_{0})$ when defining the phase space,
then this will correspond to $q=v\lambda+q(0)$ parametrization.
In this respect, for $\boldsymbol{H}=v(p_{1}-p_{1}(0))$ one has the
interpretation of the constant $p_{1}(0)$ as $\phi_{0}$ in the Lagrangian
formalism and one can see that $L=v\phi_{0}=vp_{1}(0)$.

In this spirit of reasoning, one sees that $p_{0}$ will correspond
to backward coordinate time motion when one considers $\boldsymbol{H}=p_{0}-E$,
i.e for $f=t$ one has: 
\begin{equation}
\frac{df}{d\lambda}=\left\llbracket f,\boldsymbol{H}\right\rrbracket \rightarrow\frac{dt}{d\lambda}=\left\llbracket t,p_{0}\right\rrbracket \rightarrow\frac{dt}{d\lambda}=-1
\end{equation}
and the energy $p_{0}$ is conserved, which is consistent with the
choice of a manifold structure determined by $\boldsymbol{H}=p_{0}-E=0$.

From the above discussion, one can conclude that the extended Hamiltonian
$\boldsymbol{H}$ can also reflect space transformations back and
forth along a spatial coordinate. Therefore, $\boldsymbol{H}$ should
be viewed as a generator of transformations along a path 
$\boldsymbol{H}=0$ reflects the laboratory coordinate expression
of the path as viewed by the observer and gives the relationship between
the laboratory coordinate $q_{1}$ and the corresponding momentum
$p_{1}$ along the space-like path. When the path is a process, thus
time-like related curve, then $\boldsymbol{H}$ is the extended Hamiltonian
describing the relationship between the time coordinate $t$ and the
energy of the process $E=p_{0}$ as seen by the observer.

Going back to the one spatial coordinate case, the corresponding quantum
picture now is based on $\boldsymbol{H}=p_{1}-p_{1}(0)$ and gives:
\begin{equation}
\boldsymbol{H}=p_{1}-\phi(q)\rightarrow\boldsymbol{\hat{H}}\psi(t)=0\Rightarrow-i\hbar\frac{\partial\psi}{\partial q}=\phi(q)\psi
\end{equation}
where the wave function solutions $\psi(q)$ are given by: 
\begin{equation}
\psi(q)=\frac{1}{\mathcal{N}}\exp\left[\frac{i}{\hbar}\left(\intop_{0}^{q<\delta}\phi(q)dq
+p_{1}\intop_{\delta}^{q\gg\delta}dq\right)\right]
\label{wave function psi(q)}
\end{equation}

For spatial coordinate interval of the process $\Delta q\gg\delta$
such that the momentum $p_{1}$ is conserved, as discussed earlier
in the Lagrangian formulation of this system (\ref{eq:Conservation of momentum-Statistically}),
this is the familiar plane wave with normalization factor $\mathcal{N}=1$
if one switches to $q=vt$. However, for fluctuation of $\phi(q)$ at
short length scale $\Delta q\apprle\delta$ that does not show momentum
conservation for the process then the wave function is related to
the integral of $\phi(q)$ and the normalization $\mathcal{N}$ may
depend on the size of $\delta$ and the structure of the relevant Hilbert
space. Again the inner product in the space of solutions that turns
it into a Hilbert space could be tricky, but a running average may
be useful. However, for the plane waves one can use the standard inner
product where $\Delta$ would be a sufficiently long observational
window for the process such that momentum is conserved and thus slight
variations in the window size $\Delta$ are producing consistent
results for the structure of the Hilbert space: 
\[
\left\langle \Psi|\Phi\right\rangle _{\Delta}=\frac{1}{\Delta}\intop\Psi^{*}\Phi dq
\]

Unlike the one-time coordinate case, here one does not have any limitation
on the sign of the linear momentum and if one views the process as a
moving particle with velocity $v$ then the complex conjugated wave
function $\psi(q)^{*}$ would correspond to a particle moving with
opposite momentum or equivalently with opposite direction of the velocity.
The expression (\ref{wave function psi(q)}) 
shows that the complex conjugated wave function $\psi(q)^{*}$ 
should, therefore, be viewed as the wave function for the 
directionally reversed process of the original process.

One can also construct the extended Hamiltonian $\boldsymbol{H}$
for the proper-length parameterization or motion along a curved path
by looking at $\boldsymbol{H}=(p_{1}/\phi(q_{1}))-\chi(q_{1})$ which
gives us the proper-length relationship $dl=d\lambda=\phi(q_{1})dq_{1}$
and $\chi(q_{1})$ should be such that the Hamiltonian constraint
$\boldsymbol{H}=0$ gives $p_{1}=\phi(q_{1})$ as before. Thus, $\chi(q_{1})=1$
finally gives us the $\boldsymbol{H}$ for proper-length parameterization:
\[
\boldsymbol{H}=\frac{p_{1}}{\phi(q_{1})}-1
\]

\begin{equation}
\frac{df}{d\lambda}=\left\llbracket f,\boldsymbol{H}\right\rrbracket \rightarrow\frac{dq_{1}}{d\lambda}=\left\llbracket q_{1},\frac{p_{1}}{\phi(q_{1})}-1\right\rrbracket \rightarrow\frac{dq_{1}}{d\lambda}=\frac{1}{\phi(q_{1})}\left\llbracket q_{1},p_{1}\right\rrbracket =\frac{1}{\phi(q_{1})}
\end{equation}
\begin{equation}
\frac{df}{d\lambda}=\left\llbracket f,\boldsymbol{H}\right\rrbracket \rightarrow\frac{dp_{1}}{dl}=p_{1}\left\llbracket p_{1},\frac{1}{\phi(q_{1})}\right\rrbracket =p_{1}\frac{1}{\phi(q_{1})^{2}}\frac{\partial\phi(q_{1})}{\partial q_{1}}=\frac{\partial\ln\phi(q_{1})}{\partial q_{1}}=\frac{d\phi(q_{1})}{dl}
\end{equation}
This corresponds to the previous result obtained in the Lagrangian
formalism.

The corresponding quantum picture now is very similar to the proper-time
quantization with additional term in the Schr\"odinger like equation:
\begin{equation}
\boldsymbol{H}=\frac{1}{2}\left(\frac{1}{\phi(q)}p_{1}+p_{1}\frac{1}{\phi(q)}\right)-1\rightarrow\boldsymbol{\hat{H}}\psi(t)=0
\end{equation}
\begin{equation}
\frac{1}{2}\left(\frac{1}{\phi(q)}\hat{p}_{1}\psi(q)+\hat{p}_{1}\frac{\psi(q)}{\phi(q)}\right)=\psi(q)
\end{equation}
\begin{equation}
-\frac{i\hbar}{2}\frac{1}{\phi(q)}\left(\frac{\partial\psi}{\partial q}+\phi(q)\frac{\partial}{\partial q}\left(\frac{\psi(q)}{\phi(q)}\right)\right)=\psi(t)
\end{equation}
\begin{equation}
-i\hbar\frac{\partial\psi}{\partial q}=\left[\phi(q)-\frac{i\hbar}{2}\left(\frac{\partial\ln\phi(q)}{\partial q}\right)\right]\psi(q)
\end{equation}

Therefore, the amplitude of the original wave function will be modulated
now by an additional factor $\varrho(q)$ satisfying: 
\begin{equation}
\frac{\partial\varrho(q)}{\partial q}=\frac{1}{2}\left(\frac{\partial\ln\phi(q)}{\partial q}\right)\varrho(q)
\end{equation}
This factor, $\varrho(q)=\sqrt{\phi(q)}$, will now become the main
part of the wave function if one views the system in the center of mass
frame where $p_{1}=0$. 
\vspace{6pt}
\begin{equation}
\psi(q)=\frac{1}{\mathcal{N}}\sqrt{\phi(q)}
\exp\left[\frac{i}{\hbar}\left(\intop_{0}^{q<\delta}\phi(q)dq
+p_{1}\intop_{\delta}^{q\gg\delta}dq\right)\right]
\rightarrow\frac{1}{\mathcal{N}}\sqrt{\phi(q)}
\end{equation}

In general, the normalizability of the wave-function implies positivity of $p_1>0$ 
as in the previous case of the $p_0$ in the proper-time parametrization. However,
the directionality is encoded in the sign of the phase factor and whether one is looking at 
$\psi$ or its complex conjugate $\psi^{*}$. Thus, for conserved non-zero momentum $p_1>0$ 
the normalization becomes~$\mathcal{N}^2=p_1$: 
\[
\left\Vert \psi\right\Vert ^{2}=\left\langle \psi|\psi\right\rangle _{\Delta}
=\frac{1}{\mathcal{N}^{2}\Delta}\intop_{0}^{\Delta>>\delta}\phi(q)dq
\underset{\Delta\rightarrow\infty}{\longrightarrow}\frac{p_1}{\mathcal{N}^{2}}.
\]

However, in the center of mass frame where $p_{1}=0$ this changes the wave-function normalization 
$\mathcal{N}$ to be related to the details of the quantum fluctuations of $\phi(q)$ since 
$\delta<<\Delta\rightarrow\infty$ leads to:

\[
\left\Vert \psi\right\Vert ^{2}=\left\langle \psi|\psi\right\rangle _{\Delta}=
\frac{1}{\mathcal{N}^{2}\Delta}\intop_{0}^{\Delta\gg\delta}\phi(q)dq\underset{\Delta\rightarrow\infty}{\longrightarrow}0.
\]

Thus the effects of the quantum phenomenon disappear when
the system is viewed at coarse-grain scale $\Delta\gg\delta$.
This may indicate that the inner product in the Hilbert space may
have to be redefined:
\[
\left\langle \Psi|\Phi\right\rangle =\intop\Psi^{*}\Phi dq
\]
\textbf{Now the normalizability of the wave function is related to
the usual spatial localization of the physically relevant states} 
$\psi$ that was modulated by the factor $\sqrt{\phi}$.

\[
\left\Vert \psi\right\Vert ^{2}=\left\langle \psi|\psi\right\rangle=
\frac{1}{\mathcal{N}^{2}}\intop_{0}^{\delta}\phi(q)dq.
\]

\subsubsection{The Notion of Time Reversal}

In the discussion above, we have shown that the meaning of the process
time parameterization $\lambda$ is intimately related to the choice
of the Hamiltonian constraint $\boldsymbol{H}$ as expressed in the
laboratory. Changing $\boldsymbol{H}$ to its negative $\boldsymbol{H}\rightarrow-\boldsymbol{H}$
does not change the phase space determined by the Hamiltonian constraint
$\boldsymbol{H}=0$, but changes the choice of parametrization $\lambda$
to $\xi$ that are now time reversal to each other $d\xi=-d\lambda$.
One can see this by comparing the evolution equations of the coordinate time $t$: 
\begin{equation}
\frac{dt}{d\lambda}=\left\llbracket t,\boldsymbol{H}\right\rrbracket \rightarrow\frac{dt}{d\xi}=\left\llbracket t,(-\boldsymbol{H)}\right\rrbracket =-\frac{dt}{d\lambda}
\end{equation}
Thus, if one considers 
\begin{equation}
\boldsymbol{H}=p_{0}-\phi(t)
\end{equation}
in the earlier example above, then one would have: 
\begin{equation}
\frac{dt}{d\xi}=\left\llbracket t,\boldsymbol{H}\right\rrbracket \rightarrow\frac{dt}{d\xi}=\left\llbracket t,p_{0}-\phi(t)\right\rrbracket =-1
\end{equation}
from where one can deduce that $d\xi=-dt$. If one observed that the energy
$E=p_{0}$ did not change during the process then this will correspond
to a time reversal process. For example, if there are two ``identical''
clocks one in the laboratory and the other outside and one observes
and compare the time from both. Then, one can conclude that one clock
is running backwards. This way, it will be possible for models based
on reparametrization invariance formalism to have 
time reversal as a symmetry 
along with the common arrow of time due to the positivity of the energy (the rest
mass of the observers).

\subsection{The Meaning of $\lambda$ and $\bfifH$ in the Extended Phase-Space}

From the previous discussion, we understand that the phase-space momentum
coordinates $p_{i}$ can be considered as generators of forward motion
along the corresponding coordinates $q_{i}$, while the time and energy
coordinate stand out in that $p_{0}$ will correspond to backward
coordinate time transformation. In a similar way, the extended Hamiltonian
$\boldsymbol{H}$ defines the evolution of a system's observables
$f$ along a process parametrized by $\lambda$. In the observer's
coordinate frame, $\boldsymbol{H}$ defines the relevant phase-space
via $\boldsymbol{H}=0$ along with equations that tell the observer
how the process will unfold from one stage (state), determined by
a point in the phase-space, to the nearby stage (state)---another point in the phase space. This is different from the Lagrangian
formulation where the configuration space $M$ and its co-tangential
space $TM$ that contains the coordinates and the velocities have
to be ``predetermined'' in a way that has nothing to do with the
Lagrangian $L$. The Lagrangian, however, tells how the process
should be embedded in the tangential space $TM$ by using the Euler--Lagrange
equations of motion expressed in a specific laboratory coordinate
frame. The phase-space, in this case, is determined by the initial
conditions and it is expected to be a sub-manifold of $TM$ upon the
evolution using the Euler--Lagrange equations. In the laboratory coordinate
frame, the choice $\lambda=t$ is the natural first choice for the
process parametrization. However, upon investigation of the system
in the Lagrangian formulation one may arrive at the notion of a proper-time
$\tau$ that may be a more useful choice of parametrization of a process
that should be detached from the choice of a laboratory coordinate
frame in the sense that this is the unique laboratory frame where
all the special velocities are zero and the time-speed $u$ is 1. 
That is, in an arbitrary laboratory frame the various momenta
are determined from $p_{\mu}=\partial L/\partial v^{\mu}$ and evolve
according to Euler--Lagrange equations $dp_{\mu}/d\lambda=\partial L/\partial x^{\mu}$,
but there is the unique co-moving frame where $v^{0}=dx^{0}/d\lambda=dt/d\lambda=1$
and $v^{i}=dq^{i}/d\lambda=0$. Then, for homogeneous Lagrangians of
first-order the phase space should be determined by an additional
requirement such as parallel transport that conserves the norm of
the vectors ($dl/d\lambda=0$).

In the extended Hamiltonian framework, the phase-space is determined
from $\boldsymbol{H}=0$ and the evolution of the coordinates and
momenta are governed by the evolution equation via the extended Poisson
bracket $df/d\lambda=\left\llbracket f,\boldsymbol{H}\right\rrbracket $
the specific choice of $\boldsymbol{H}$ then tells us the details
about the coordinate frame where the observer is studying the process.
The reparametrization-invariance was explicit in the Lagrangian framework
due to the use of first-order homogeneous Lagrangians in the velocities.
In the extended Hamiltonian formulation this is somehow encoded in
the extended Hamiltonian $\boldsymbol{H}$ and the structure of the
phase-space determined from $\boldsymbol{H}=0$. To understand
how the extended Hamiltonian should change when one changes the choice
of parametrization one can consider the extended Poisson bracket evolution
equation for two different parametrizations that are related by $\lambda(\xi)$:
\[
\frac{df}{d\lambda}=\left\llbracket f,\boldsymbol{H}_{\lambda}\right\rrbracket \rightarrow\left(\frac{d\xi}{d\lambda}\right)\frac{df}{d\xi}
=\left\llbracket f,\boldsymbol{H}_{\lambda}\right\rrbracket \rightarrow\left(\frac{d\xi}{d\lambda}\right)\left\llbracket f,\boldsymbol{H}_{\xi}\right\rrbracket 
=\left\llbracket f,\boldsymbol{H}_{\lambda}\right\rrbracket 
\]

This can be satisfied if $d\xi\boldsymbol{H}_{\xi}=d\lambda\boldsymbol{H}_{\lambda}+d\lambda I$
where $I$ is an integral of the process $(\left\llbracket I,\boldsymbol{H}\right\rrbracket =0)$
such that $I=0$ over the phase space determined by $\boldsymbol{H}=0$.
To illustrate this let us consider $\boldsymbol{H}_{t}=\phi(t)-p_{0}$
and $\boldsymbol{H}_{\tau}=1-\frac{p_{0}}{\phi(t)}$. From the specific
expressions one can see that $\boldsymbol{H}_{t}=\phi(t)\boldsymbol{H}_{\tau}$
thus $d\tau\boldsymbol{H}_{\tau}=dt\boldsymbol{H}_{t}$ and therefore
$d\tau=\phi(t)dt$, which is the usual definition of the relationship
of the proper-time to the coordinate time.

If one applies this framework to a moving particle with constant velocity
$v$ along the spatial coordinate $q$ one has $q=vt+q(0)$ where
$t$ is the new parametrization. Therefore, $\boldsymbol{H}_{q}=p_{1}-p_{1}(0)$
should be related to $\boldsymbol{H}_{t}=v\boldsymbol{H}_{q}+vI=v(p_{1}-p_{1}(0))+vI$.
The question now is what is the integral of motion $I$? To find it,
one should realize that the configuration space now is two-dimensional
$(t,q)$ and the phase-space then will also include $(p_{0},p_{1})$.
Therefore, $I$ can be determined from the evolution of the equation
for the coordinate $t$ and from the requirement that $I$ is an integral
of motion: 
\[
\frac{df}{dt}=\left\llbracket f,\boldsymbol{H}_{t}\right\rrbracket \rightarrow\frac{dt}{dt}=\left\llbracket t,\boldsymbol{H}_{t}\right\rrbracket =\left\llbracket t,vI\right\rrbracket =1\Rightarrow I=-\frac{p_{0}}{v}
\]

This way the corresponding general expression for $\boldsymbol{H}_{t}$
becomes: 
\[
\boldsymbol{H}_{t}=v(p_{1}-p_{1}(0))-(p_{0}-p_{0}(0))\rightarrow\overrightarrow{v}\cdot\overrightarrow{p}-p_{0}+E
\]

Although this is the physically more relevant system to study due
to its possibility to include at least one spatial coordinate and
the necessary one-time-coordinate within a Minkowski space-time, it
is beyond the scope of the paper which is to analyze the simplest
reparametrization-invariant one-dimensional systems for
physically relevant consequences and to understand the meaning of
the reparametrization parameter $\lambda$. 

Based on the examples and the discussions above, one can conclude
that the role of the reparametrization parameter $\lambda$ is of a
placeholder parameter that is to be clarified after a specific choice
of the expression for $\boldsymbol{H}$. However, the usual dynamic
time-like meaning of $\lambda$ is often associated with the expression
for $\boldsymbol{H}$ that defines the whole phase-space or Hilbert
space of the system either via $\boldsymbol{H=0}$ or via the expression
for $\boldsymbol{\hat{H}\psi=0}$.

\section{Conclusions and Discussion \label{sec:Conclusions-and-Discussions}}

Following the main motivation on the importance of reparametrization-invariant models, 
we have studied the meaning and the roles of the parameter $\lambda$ 
for the simplest reparametrization-invariant system in one-dimension as well as the  physically relevant example of the relativistic particle in any dimensions.

In the process, the extended Hamiltonian formulation discussed 
was a Lorentz-invariant, and in general, which naturally
leads to the standard Schr\"odinger equation from Quantum Mechanics.
The superposition principle, which is the bedrock of Quantum Mechanics and
is a natural property of the Hilbert space defined via the Hamiltonian constraint
$\boldsymbol{\hat{H}}\Psi=0$.
From the examples studied, one can conclude that the proper-length and the
proper-time are uniquely identified as the parameterizations where
the corresponding Lagrangian becomes a constant of motion with its
value equal to 1. In the corresponding extended Hamiltonian formulation,
the corresponding extended Hamiltonian $\boldsymbol{H}$ is easily
identifiable in the coordinate $t$ parametrization. While we have
shown and confirmed the corresponding expression for the extended
Hamiltonian $\boldsymbol{H}$ in the proper-time parametrization $\tau$,
it is not clear how to identify the functional form of $\boldsymbol{H}$
in more general $n$-dimensional~systems. 

In the case of the relativistic particle in gravitational field, however,
a Hamiltonian that possesses the key properties of non-relativistic 
system in proper-time parametrization has been studied in Appendix \ref{Appendix}.

In general, the connection between the explicit form of the extended Hamiltonian $\boldsymbol{H}$
and the meaning of the parameter $\lambda$ has been illustrated clearly.
The quantum mechanical equivalent of such systems has been studied
and in the coordinate $t$ parametrization, the usual plane wave has
been recovered. An interesting result has emerged from the study of
the system using the extended Hamiltonian $\boldsymbol{H}$ for the
proper-time parametrization. The wave function now is modulated by
a field $\phi(t)$ and in the limit of energy conservation on the
macroscopic scale, the energy is forced to be positive in order to
have a normalizable wave function. 

In the case of the relativistic particle in a gravitational field, 
the field $\phi(t)$ has been identified with the relativistic factor $\gamma^{-1}$, which 
is connected to the norm of the quantum mechanical state and thus has to be positive. 
Furthermore, for a weak time-dependent gravitational field,
the energy of the particle receives a factor proportional to the rate of 
change of the gravitational field.

This implies the positivity of the rest mass when the field fluctuation can be neglected. 
Similarly, the coordinate distance $q$ recovers the
familiar plane wave with conserved momentum at macroscopic scale and
in the proper-length parametrization the wave function
now is modulated by the field $\phi(q)$, which should have localizable quantum fluctuations
in order to be normalizable. The normalizability of the wave function requires positivity of 
the energy and momentum variables while the directionality is now encoded  
in the phase factor of the quantum mechanical wave-function $\psi$ and its
complex conjugated $\psi^{*}$. 
Models based on reparametrization-invariance are likely to have 
time reversal as a symmetry along with the common arrow of time 
due to the positivity of the rest mass of the particles. 
The next steps in this study on the reparametrization-invariant models 
is to follow the above procedures and to apply them to the relativistic particle in a gravitational field,
then to extend particles with spin and to compare the results with the Dirac's formalism, 
as well as to well-known string theory models.

Some alternative directions for research are also related to explorations of the applicability of the 
extended Hamiltonian framework to the Born Reciprocity and Reciprocal Relativity (\cite{STQEmergenceBook}, Chapter 9),
and to seek appropriate non-commutative symplectic algebra~(\cite{STQEmergenceBook}, Chapter 1) extension,
as well as Quantized Fields \'a la Clifford (\cite{STQEmergenceBook}, Chapter 23), which is
in one of the original research directions \cite{VGG2002Varna,VGG2003}.

\vspace{6pt}

\acknowledgments{A.M. expresses his gratitude to his wife for her patience and support.
V.G. is extremely grateful to his wife and daughters for their understanding and family support  during the various stages of the research presented.}

\authorcontributions{
Conceptualization, V.G.; 
validation, V.G. and A.M.; formal analysis, V.G. and A.M.; 
writing---original draft preparation, V.G..; writing---review and editing, V.G. and A.M.; 
All authors have read and agreed to the published version of the manuscript.} 

\funding{ This research received no external funding. } 

\institutionalreview{Not applicable. } 

\informedconsent{Not applicable. } 

\dataavailability{Not applicable. } 

\conflictsofinterest{The authors declare no conflict of interest.} 

\appendixtitles{yes} 
\appendixstart
\appendix

\section{The Relativistic Particle}\label{Appendix}

In this section the above formalism is illustrated for the case of the relativistic spinless point particle 
in the presence of the electromagnetic and gravitational background fields.
For the strictly relativistic formalism and with more details, one can consult \cite{Deriglazov2016}.
For treatment of  particles with spin via the use of singular Lagrangians see 
\cite{2014NuPhB.885....1D,2014SIGMA..10..012D,2013arXiv1312.6247D,1999MPLA...14..709D}.
\noindent
The action $ a=\int L(x,\upsilon)d\lambda$ with (dimensionless) velocity
$\upsilon^{\mu}=dx^{\mu}/d\lambda$ is reparametrization invariant
when the following relativistic particle Lagrangian is considered:
\[
L=qA_{\mu}\upsilon{}^{\mu}-mc^{2}\sqrt{-g_{\mu\nu}\upsilon^{\mu}\upsilon^{\nu}}.
\]
Notice that in this section we have chosen a different metric signature $(-+++)$ that is more practical 
since it avoids sign changes when going from covariant to contravariant spacial components.
The corresponding Euler--Lagrange equations are:
\[
\pi_{\mu}=\frac{\partial L}{\partial\upsilon^{\mu}}=qA_{\mu}
+mc^{2}\frac{g_{\mu\nu}\upsilon^{\nu}}{\sqrt{-g_{\mu\nu}\upsilon^{\mu}\upsilon^{\nu}}},
\]
\[
\frac{d\pi_{\rho}}{d\lambda}=\frac{\partial L}{\partial x^{\rho}}=qA_{\mu,\rho}\upsilon^{\mu}
+\frac{mc^{2}}{2}\frac{g_{\mu\nu,\rho}\upsilon^{\mu}\upsilon^{\nu}}{\sqrt{-g_{\mu\nu}\upsilon^{\mu}\upsilon^{\nu}}}
\]

\subsection{Coordinate-Time Parametrization}

The choice of coordinate-time parametrization $(d\lambda=dt,\:dx^{0}=cdt)$
breaks explicitly the reparametrization invariance and results in
dimension-full velocity $v^{\mu}=(c,\overrightarrow{v})$. In what
follows $\vec{v}=(d\overrightarrow{x}/dt)$ denotes the usual three-dimensional
velocity and $v^{0}=dx^{0}/dt=c$. Furthermore, $v^{2}=\vec{v\cdot}\vec{v}=g_{ij}v^{i}v^{j}$
is considered implicitly $x$-dependent due to the metric $g_{ij}(x)$.
Notice that although the metric signature is $(-+++)$ when $g_{00}$ 
is written explicitly it will be considered to be positive $|g_{00}|=-g_{00}$.

\subsubsection{Lagrangian Formulation}

The Lagrangian and the corresponding Euler--Lagrange Equations are:
\[
L=qcA_{0}+qA_{i}v^{i}-mc^{2}\sqrt{g_{00}
-\frac{\vec{v\cdot}\vec{v}}{c^{2}}}=qcA_{0}+qA_{i}v^{i}-mc^{2}\gamma^{-1}\:,
\]

\[
\pi_{i}=\frac{\partial L}{\partial v^{i}}=qA_{i}+m\frac{v_{i}}{\sqrt{g_{00}
-\frac{\vec{v\cdot}\vec{v}}{c^{2}}}}=qA_{i}+m\gamma v_{i}\:,
\]
\[
\frac{d\pi_{i}}{dt}=\frac{\partial L}{\partial x^{i}}=qcA_{0,i}+qA_{j,i}v^{j}+mc^{2}\gamma^{-2}\gamma_{,i}\:,
\]
by setting $\gamma=c/\sqrt{-g_{\mu\nu}v^{\mu}v^{\nu}}=1/\sqrt{g_{00}-\frac{\vec{v\cdot}\vec{v}}{c^{2}}}$ and
by using the kinematical momentum $p_{i}=m\gamma v_{i}$ one has:

\[
\frac{d\pi_{i}}{dt}=\frac{d}{dt}\left(qA_{i}+p_{i}\right)=qcA_{0,i}+qA_{j,i}v^{j}+mc^{2}\gamma^{-2}\gamma_{,i}\:.
\]
Usually, however, one is interested in the rate of change of the kinematic linear momentum~$p_{i}$:

\[
\frac{dp_{i}}{dt}=qcA_{0}{}_{,i}+q\left(A_{j,i}-A_{i,j}\right)v^{j}-qcA_{i,0}+mc^{2}\gamma^{-2}\gamma_{,i}\:,
\]

\[
\frac{dp_{i}}{dt}=qcF_{0i}+q\overrightarrow{e_{i}}\cdot F\cdot\overrightarrow{v}+mc^{2}\gamma^{-2}\gamma_{,i}\:,
\]
\[
\frac{dp_{i}}{dt}=q\overrightarrow{e_{i}}\cdot\overrightarrow{E}
+q\overrightarrow{e_{i}}\cdot\overrightarrow{v}\times\overrightarrow{B}+mc^{2}\gamma^{-2}\gamma_{,i}\:.
\]
where the relation between the Faraday tensor $F_{\mu\nu}=\partial_{\mu}A_{\nu}-\partial_{\nu}A_{\mu}$
and the electric $E_{i}=cF_{0i}$ and magnetic fields $B_{i}=\frac{1}{2}\epsilon_{ijk}F^{jk}$
have be utilized. 

Furthermore, since $cd\tau=ds=\sqrt{-g_{\mu\nu}dx^{\mu}dx^{\nu}}$
then $ds/d\lambda=cd\tau/dt$ and thus\linebreak $\gamma d\tau=dt$ with $\gamma=c/\sqrt{-g_{\mu\nu}v^{\mu}v^{\nu}}$.
Then from the energy-momentum constraint one has that $(mc)^{2}=(p^{0})^{2}-\overrightarrow{p}{}^{2}$,
which when combined with $p^{2}=m^{2}\gamma^{2}v^{2}$ and $p^{0}=m\gamma c=E/c$,
gives $(mc)^{2}=(p^{0})^{2}(g_{00}-p^{2}/(p^{0})^{2})=(mc)^{2}\gamma^{2}(g_{00}-p^{2}/(p^{0})^{2})$
thus $\gamma=(g_{00}-p^{2}/(p^{0})^{2})^{-1/2}=p^{0}/\sqrt{-g_{\mu\nu}p^{\mu}p^{\nu}}$
along with $p_{0}/mc=\gamma$, which shows that $p_{0}$ is an integral
of the motion whenever $\gamma$ is an integral of the motion. Notice
that the energy-momentum constraint, the Lorentz invariance, and the
kinematic momentum expression $p_{\mu}=m\gamma v_{\mu}$ are consistently
imposed with respect to the coordinate-time parametrization $t$ and
the four-velocity is $v_{\mu}=(c,\overrightarrow{v})$.

The Hamiltonian function in the tangential space is non-zero since
the reparametrization invariance has been removed by the choice of
coordinate-time parametrization. Thus, $h(x,v)$ is related to the
energy of the system:

\[
h=\pi_{i}v^{i}-L=\left(qA_{i}+m\gamma v_{i}\right)v^{i}-\left(qcA_{0}+qA_{i}v^{i}-mc^{2}\gamma^{-1}\right),
\]
\[
h=-qcA_{0}+m\gamma c^{2}\left(v^{2}/c^{2}+\gamma^{-2}\right)=-qcA_{0}
+m\gamma c^{2}\left(\left(\frac{v^{2}}{c^{2}}\right)+\left(g_{00}-\frac{v^{2}}{c^{2}}\right)\right),
\]
\begin{equation}
h=-qcA_{0}+m\gamma c^{2}g_{00}.\label{eq:h(x,v)}
\end{equation}

In the limit of slow motion, this is the usual non-relativistic expression:
\begin{equation}
h=-qcA_{0}+mc^{2}\sqrt{g_{00}}+\frac{1}{2}\frac{m}{\sqrt{g_{00}}}\vec{v\cdot}\vec{v}
=-qcA_{0}+mc^{2}-mU+\frac{p^{2}}{2m}.\label{eq:h(x,p)-newtonian-limit}
\end{equation}

Notice that it contains only electrostatic and gravitational energy
terms and, due to the choices made, $g_{00}$ is positive ($g_{00}=(1-2U(x)/c^{2})$). 

\subsubsection{Hamiltonian Formulation}

In general, for the Hamiltonian formulation one has to use the expression
of $h(x,\pi)$ in the $\pi$-space by using $v^{i}=p^{i}/(m\gamma)$ and
$p_{i}=\pi_{i}-qA_{i}$. The Hamiltonian function of the system is then:

\[
h=\pi_{i}v^{i}-L=\frac{1}{m\gamma}\pi_{i}\left(\pi^{i}-qA^{i}\right)
-\left(qcA_{0}+\frac{1}{m\gamma}qA^{i}\left(\pi_{i}-qA_{i}\right)-mc^{2}\gamma^{-1}\right),
\]

\begin{equation}
h=\frac{1}{m\gamma}\left(\pi_{i}-qA_{i}\right)\left(\pi^{i}-qA^{i}\right)-qcA_{0}+mc^{2}\gamma^{-1}\:.\label{eq:h-in-T*M}
\end{equation}
Working with this Hamiltonian may results in an extra factor of {two},
if one is not taking into account the derivative of the term $mc^{2}\gamma^{-1}$:
\[
v^{i}=\frac{dx^{i}}{dt}=\{x^{i},h\}=\frac{\partial h}{\partial\pi_{i}}=
\frac{{2}}{m\gamma}\left(\pi^{i}-qA^{i}\right)
+\left(\frac{1}{m}\left(\pi_{i}-qA_{i}\right)\left(\pi^{i}-qA^{i}\right)
+mc^{2}\right)\frac{\partial\gamma^{-1}}{\partial\pi_{i}}.
\]
Upon using $p_{i}=\pi_{i}-qA_{i}$ along with $p_{0}=m\gamma c$ and
$\gamma^{-1}=\sqrt{g_{00}-p^{2}/(p^{0})^{2}}$, which implies 
$\frac{\partial\gamma^{-1}}{\partial p_{i}}=-\gamma p^{i}/(p^{0})^{2}$, 
one gets the correct equations by taking into account the non-relativistic
limit $m^{2}c^{2}\gg p_{i}p^{i}$:
\[
v^{i}=\frac{{2}}{m\gamma}p^{i}-\left(\frac{1}{m}p_{i}p^{i}
+mc^{2}\right)\frac{\gamma p^{i}}{(p^{0})^{2}}=\frac{{2}}{m\gamma}p^{i}
-\left(\frac{1}{m}p_{i}p^{i}+mc^{2}\right)\frac{\gamma p^{i}}{(m\gamma c)^{2}},
\]
\[
v^{i}=\frac{1}{m\gamma}p^{i}\left[1-(\frac{1}{m}p_{i}p^{i})/(mc^{2})\right]\approx\frac{1}{m\gamma}p^{i}.
\]
While the rate of change of $\pi_{i}$ is given by:
\[
\frac{d\pi_{i}}{dt}=\{\pi_{i},h\}=-\frac{\partial h}{\partial x^{i}}=qcA_{0,i}
+\frac{1}{m\gamma}p^{j}p^{k}g_{jk,i}+\left(\frac{1}{m}p_{i}p^{i}+mc^{2}\right)\gamma^{-2}\gamma_{,i}\:.
\]
The relationship $p^{0}=m\gamma c=E/c$ can be reinforced via the
extended Hamiltonian approach where $\boldsymbol{H}=0$ for the following $\boldsymbol{H}$:
\begin{equation}
\boldsymbol{H}=h-cp^{0}\:.\label{eq:H4t-time}
\end{equation}
However, starting with (\ref{eq:h-in-T*M}) it is not clear how to arrive at at 
$p^{\mu}=m\gamma v^{\mu}$ and $\gamma^{-1}=\sqrt{g_{00}-p^{2}/(p^{0})^{2}}$
unless one considers (\ref{eq:h(x,v)}) and requires that the two
expressions for $h$ to always be equal, then one will reproduce the
energy-momentum constraint: 
\[
mc^{2}\gamma g_{00}=\frac{1}{m\gamma}p^{2}+mc^{2}\gamma^{-1}\Rightarrow(p^{0})^{2}g_{00}-p^{2}=(mc)^{2}.
\]
Then, by using $p^{0}=m\gamma c=E/c$ and $p^{i}=m\gamma v^{i}$,
one can be recover $\gamma$ and so on.

The evolution of $x^{0}$ and $p^{0}$ in the extended Hamiltonian
framework will then be:
\[
\frac{dx^{0}}{d\lambda}=\left\llbracket x^{0},\boldsymbol{H}\right\rrbracket 
=\left\llbracket x^{0},-cp^{0}\right\rrbracket =c\left\llbracket x^{0},\,p_{0}\right\rrbracket =c.
\]
Therefore, because $x^{0}=ct$, in this case one has $d\lambda=dt$;
thus, this choice of $\boldsymbol{H}$ corresponds to the coordinate-time parametrization.
Finally, since $h$ has no explicit time dependence, then the energy
$cp^{0}=E$ is an integral of the motion whenever $\gamma$ is an
an integral of the motion and the metric is static:
\[
\frac{dp^{0}}{d\lambda}=\left\llbracket p^{0},\boldsymbol{H}\right\rrbracket 
=\left\llbracket p^{0},h\right\rrbracket =-\frac{\partial h}{\partial x^{0}}
=\left(\frac{1}{m}p^{2}+mc^{2}\right)\gamma^{-2}\gamma_{,0}-\frac{1}{m\gamma}p^{i}p^{j}g_{ij,0}.
\]

Quantization can proceed by using either (\ref{eq:h(x,p)-newtonian-limit})
or (\ref{eq:h-in-T*M}) for the non-relativistic Schrodinger equation.
Based on the corresponding Schrodinger's equation, as discussed in
the main text, the standard Newtonian-time quantum mechanics and its interpretations
follows naturally. As discussed earlier, in the coordinate-time parametrization,
the Newtonian-time quantum mechanics is trivially consistent with extended
Hamiltonian formalism using $\boldsymbol{H}=h-cp^{0}=0$ along with the extended
Poisson bracket. 

An interesting question is what parametrization, within the extended
Hamiltonian formalism, would $\boldsymbol{H}=1-cp_{0}/h$ correspond to, and what
is the extended Hamiltonian that corresponds to the proper-time parametrization?
Before we consider these options let us go over the proper-time parametrization
case in the Lagrangian formulation.

\subsection{Proper-Time Parametrization}

\subsubsection{Lagrangian Formulation }

In proper-time parametrization 
($d\lambda=d\tau=ds/c,\:u^{\mu}=dx^{\mu}/d\tau, \:\sqrt{-g_{\mu\nu}u^{\mu}u^{\nu}}=c$):
\[
L=qA_{\mu}u^{\mu}-mc\sqrt{-g_{\mu\nu}u^{\mu}u^{\nu}}
\]
\[
\pi_{\mu}=\frac{\partial L}{\partial u^{\mu}}=qA_{\mu}
+mc\frac{g_{\mu\nu}u^{\nu}}{\sqrt{-g_{\mu\nu}u^{\mu}u^{\nu}}}=qA_{\mu}+mu_{\mu},
\]

\[
\frac{d\pi_{\rho}}{d\lambda}=\frac{\partial L}{\partial x^{\rho}}=qA_{\mu,\rho}u^{\mu}
+\frac{mc}{2}\frac{g_{\mu\nu,\rho}u^{\mu}u^{\nu}}{\sqrt{-g_{\mu\nu}u^{\mu}u^{\nu}}},
\]
\[
\frac{d}{d\lambda}(qA_{\rho}+mu_{\rho})=qA_{\mu,\rho}u^{\mu}+\frac{m}{2}g_{\mu\nu,\rho}u^{\mu}u^{\nu},
\]
\[
\frac{dp_{\rho}}{d\lambda}=q(A_{\mu,\rho}-A_{\rho,\mu})u^{\mu}+\frac{m}{2}g_{\mu\nu,\rho}u^{\mu}u^{\nu},
\]

\textls[-12]{Hamiltonian function is identically zero upon the use of the constraint
$(-g_{\mu\nu}u^{\mu}u^{\nu})=c^{2}$:}

\[
h=\pi_{\mu}u^{\mu}-L
\]

\[
h=(qA_{\mu}+mu_{\mu})u^{\mu}-qA_{\mu}u^{\mu}+mc^{2}\equiv0
\]

\subsubsection{Hamiltonian Formulation}

Hamiltonian function using the generalized momentum variables $\pi$:

\[
h=\pi_{\mu}u^{\mu}-qA_{\mu}u^{\mu}+mc^{2}=(\pi_{\mu}-qA_{\mu})u^{\mu}
+mc^{2}=\frac{1}{m}(\pi_{\mu}-qA_{\mu})(\pi^{\mu}-qA^{\mu})+mc^{2}
\]
In this form of the Hamiltonian function $h$ (requiring the zero
value as in the $TM$ space form $h(x,v)\equiv0$) implies the energy-momentum
constraint for the chosen metric signature:
\begin{equation}
h=\frac{1}{m}p_{\mu}p^{\mu}+mc^{2},\label{eq:h(p,p)}
\end{equation}
 
\[
u^{\mu}=(\pi^{\mu}-qA^{\mu})/m\Rightarrow p^{\mu}=mu^{\mu}\Rightarrow p_{\mu}p^{\mu}=-m^{2}c^{2}
\]

This effectively results in the Klein--Gordon equation $-g_{\mu\nu}p^{\mu}p^{\nu}+m^{2}c^{2}=0$.
However, the corresponding equations of motion produce an extra factor
of two due to quadratic dependence of $h$ on the momentum variables:
\[
u^{\rho}=\frac{dx^{\rho}}{d\lambda}=\{x^{\rho},h\}=\frac{\partial x^{\rho}}{\partial x^{\mu}}\frac{\partial h}{\partial\pi_{\mu}}
=\frac{\partial h}{\partial\pi_{\rho}}=\frac{{{2}}}{m}(\pi^{\rho}-qA^{\rho}),
\]
upon using the result $mu^{\rho}={2}(\pi^{\rho}-qA^{\rho})$
in the next calculations the term quadratic in the velocity gets a
factor of $m/4$ instead of $m/2$ as in the Euler--Lagrange equations:
\[
\frac{d\pi_{\rho}}{d\lambda}=\{\pi_{\rho},h\}=-\frac{\partial h}{\partial x^{\rho}}=\frac{q}{m}A_{\mu,\rho}(\pi^{\mu}-qA^{\mu})
+\frac{q}{m}(\pi_{\mu}-qA_{\mu})A_{,\rho}^{\mu}+\frac{m}{{4}}g_{\mu\nu,\rho}u^{\mu}u^{\nu},
\]
 
\[
\frac{d\pi_{\rho}}{d\lambda}=\frac{q}{{2}}A_{\mu,\rho}u^{\mu}+\frac{q}{{2}}u_{\mu}A_{,\rho}^{\mu}
+\frac{m}{{4}}g_{\mu\nu,\rho}u^{\mu}u^{\nu}
=q\left(u^{\mu}A_{\mu}\right)_{,\rho}+\frac{m}{{4}}g_{\mu\nu,\rho}u^{\mu}u^{\nu}.
\]

The issue can be resolved by using a modified Hamiltonian $\widetilde{h}=h/2$,
which is justified since $h=0$ anyway:
\[
\widetilde{h}=\frac{1}{2m}(\pi_{\mu}-qA_{\mu})(\pi^{\mu}-qA^{\mu})+\frac{m}{2}c^{2}.
\]

The corresponding equations of motion now match well the Lagrangian
formulation: 
\[
u^{\rho}=\frac{dx^{\rho}}{d\lambda}=\{x^{\rho},\widetilde{h}\}
=\frac{\partial\widetilde{h}}{\partial\pi_{\rho}}=\frac{1}{m}(\pi^{\rho}-qA^{\rho}),
\]
\[
\frac{d\pi_{\rho}}{d\lambda}=\{\pi_{\rho},\widetilde{h}\}=-\frac{\partial\widetilde{h}}{\partial x^{\rho}}
=\frac{q}{2}A_{\mu,\rho}u^{\mu}+\frac{q}{2}u_{\mu}A_{,\rho}^{\mu}+\frac{m}{2}g_{\mu\nu,\rho}u^{\mu}u^{\nu}
=\left(qA_{\mu}u^{\mu}\right)_{,\rho}+\frac{m}{2}g_{\mu\nu,\rho}u^{\mu}u^{\nu}.
\]

By looking at $dx^{0}/d\lambda$, within the expression above, one
see that $mcdt=(\pi^{0}-qA^{0})d\lambda$, which is not quite the expression
expected in the proper-time parametrization gauge where one should
have had $cd\tau=d\lambda\sqrt{-g_{\mu\nu}u^{\mu}u^{\nu}}=cdt\sqrt{-g_{\mu\nu}v^{\mu}v^{\nu}}$.
Furthermore, from the expression $dx^{0}/d\lambda$ one also has $cdt/d\lambda=u^{0}=p^{0}/m=E/(mc)$,
which should be related to $\gamma$, so one can check if $p^{0}$
is a constant of the motion since $E$ is expected to be:
\vspace{6pt}
\[
\frac{dp^{0}}{d\lambda}=\frac{d(\pi^{0}-qA^{0})}{d\lambda}=\{\pi^{0}-qA^{0},\widetilde{h}\}
=\{\pi^{0},\widetilde{h}\}-q\{A^{0},\widetilde{h}\}=\frac{d\pi^{0}}{d\lambda}
-q\frac{\partial A^{0}}{\partial x^{\mu}}\frac{\partial\widetilde{h}}{\partial\pi_{\mu}},
\]
\[
\frac{dp^{0}}{d\lambda}=\frac{d\pi_{\mu}g^{\mu0}}{d\lambda}-q\frac{\partial A^{0}}{\partial x^{\mu}}u^{\mu}
=\pi_{\mu}\frac{dg^{\mu0}}{d\lambda}+g^{\rho0}\left[\left(qA_{\mu}u^{\mu}\right)_{,\rho}
+\frac{m}{2}g_{\mu\nu,\rho}u^{\mu}u^{\nu}\right]-q\frac{dA^{0}}{d\lambda}.
\]
The above expression is quite complicated and not obviously zero,
unless the metric tensor is the Lorentz metric ($g_{\mu\nu}=\eta_{\mu\nu}$)
for a particle with a zero charge and if the solutions are considered
to be the geodesic lines with geodesically transported metric. Thus,
in the presence of general gravitational and electromagnetic fields,
it may not correspond to an integral of the motion as expected for
the energy of a system. The Hamiltonian $\widetilde{h}$ reproduces
the corresponding Euler--Lagrange equations derived in proper-time
parametrization\linebreak\textls[+12]{ ($d\lambda=d\tau=ds/c$) in the Lagrangian formalism,
guarantees the mass-shell condition}\linebreak $(-g_{\mu\nu}p^{\mu}p^{\nu}+m^{2}c^{2}=0)$
for solutions satisfying $\boldsymbol{H}=\widetilde{h}=0$, and in this respect
is related to the Klein--Gordon equation upon quantization; however,
it is not clear if this Hamiltonian formulation is really in the proper-time
parametrization of the co-moving frame of the particle since $cdt/d\lambda=(\pi^{0}-qA^{0})/m=p^{0}/m$
is not connected to $\gamma$ in a clear way.

In the co-moving frame of the particle one would expect $u^{\mu}=(c,0,0,0)$
and therefore $\pi^{\rho}=qA^{\rho}+mc\delta_{0}^{\rho}$, which gives:
\[
\frac{d\pi_{\rho}}{d\tau}=q\frac{dA_{\rho}}{d\tau}=qA_{\rho,\mu}u^{\mu}=qcA_{\rho,0},
\]
while the general equations of motion for $d\pi_{\rho}/d\tau$ results
in:
\[
\frac{d\pi_{\rho}}{d\tau}=\left(qcA_{0}\right)_{,\rho}+\frac{m}{2}c^{2}g_{00,\rho},
\]
This gives a relationship that is not guaranteed in general since
gravity and electromagnetism are independent fields:
\[
\Rightarrow qA_{\rho,0}=qA_{0,\rho}+\frac{mc}{2}g_{00,\rho}.
\]
However, given that $\:\sqrt{-g_{\mu\nu}u^{\mu}u^{\nu}}=c$, then
one concludes, using the co-moving frame of the particle where $u^{\mu}=(c,0,0,0)$,
that $g_{\mu=0,\nu=0}=-1$ and therefore $g_{00,\rho}=0$. Furthermore,
in this case the earlier expression for $dp^{0}/d\tau$ shows that
$p^{0}$ is constant of the motion since it is equal to $mc$ due
to $\pi^{\rho}=qA^{\rho}+mc\delta_{0}^{\rho}$:

\[
\frac{dp^{0}}{d\tau}=\frac{d\pi^{0}}{d\tau}
-q\frac{\partial A^{0}}{\partial x^{\mu}}\frac{\partial\widetilde{h}}{\partial\pi_{\mu}}
=\frac{d\pi^{0}}{d\tau}-q\frac{\partial A^{0}}{\partial x^{\mu}}u^{\mu}
=\frac{d\pi^{0}}{d\tau}-q\frac{dA^{0}}{d\tau}=\frac{d(mc)}{d\tau}=0.
\]

The quantum picture in this case (in the co-moving frame and based
on the Klein--Gordon equation) becomes trivial in the sense that $p^{\mu}=(E/c,0,0,0)$;
thus, the state of the system is represented by a wave function $\Psi=\exp(-\frac{i}{\hbar c}tE)$
such that $\hat{P}^{\mu}\Psi=p^{\mu}\Psi$ where $\hat{P}^{i}=-i\hbar\partial^{i}$
and $c\hat{P}^{0}=-i\hbar c\partial^{0}$, or equivalently $c\hat{P}^{0}=i\hbar\partial_{t}$
since $x^{0}=ct$. Thus, $\hat{P}^{0}\Psi=p^{0}\Psi$ with $p^{0}=E/c$.
However, this corresponds to the coordinate-time parametrization of
a co-moving observer seeing a plane-wave, which is not that much about
the proper-time parametrization of the process itself.

\subsection{Extended Hamiltonian Framework}

\subsubsection{Proper-Time Parametrization }

To have proper-time parametrization in the Extended Hamiltonian formulation one needs to have 
$ds=cd\tau=d\lambda\sqrt{-g_{\mu\nu}u^{\mu}u^{\nu}}=cdt\sqrt{-g_{\mu\nu}v^{\mu}v^{\nu}}=c\gamma^{-1}dt$,
by looking at $h$ in\linebreak (\ref{eq:h-in-T*M}), by taking $q=0$ for simplicity
since then $\pi_{\mu}=p_{\mu}$, then one can consider the following
extended Hamiltonian that provides effectively the same Hamiltonian
constraint\linebreak $\boldsymbol{H}=h+p_{0}c=0$ as before (\ref{eq:H4t-time}):
\begin{equation}
\boldsymbol{H}=(h+p_{0}c)\gamma=h_{cl}+p_{0}c\gamma
=\left(\frac{1}{2m}p_{i}p^{i}+mc^{2}\right)+p_{0}c\gamma\label{eq:H_for_tau-time}
\end{equation}
This expression reduces to $p^{0}=mc$ upon $\boldsymbol{H}=0$ at $\overrightarrow{p}=0$,
then upon $\boldsymbol{H}=0$ the non-relativistic energy follows as well since
$p_{0}=-p^{0}$ and in the co-moving frame $\gamma\approx1$. The
extended Hamiltonian formalism then shows that $\lambda$ is the proper
time parametrization upon keeping $\gamma$ constant ($\gamma\approx1$):
\[
\frac{dx^{0}}{d\lambda}=\left\llbracket x^{0},\boldsymbol{H}\right\rrbracket 
=\frac{\partial x^{0}}{\partial x^{\mu}}\frac{\partial \boldsymbol{H}}{\partial p_{\mu}}=\frac{\partial \boldsymbol{H}}{\partial p_{0}}
=c\gamma+p_{0}c\frac{\partial\gamma}{\partial p_{0}}\approx c\gamma\Rightarrow\gamma d\lambda=dt.
\]
In this case, $p_{0}$ is an integral of the motion for static spatial
metric: 
\[
\frac{dp_{0}}{d\lambda}=\left\llbracket p_{0},\boldsymbol{H}\right\rrbracket 
=-\frac{\partial p_{0}}{\partial p_{\mu}}\frac{\partial \boldsymbol{H}}{\partial x^{\mu}}
=-\frac{\partial \boldsymbol{H}}{\partial x^{0}}=\frac{1}{2m}p^{i}p^{j}(-g_{ij,0})
-p_{0}c\gamma_{,0}\approx\frac{1}{2m}p^{i}p^{j}(-g_{ij,0})\:.
\]
As expected in the instantaneous co-moving frame, the equations of
motion correspond to those generated by the classic non-relativistic
Hamiltonian $h_{cl}=p^{i}p_{i}/(2m)+mc^{2}$ since $\gamma\approx1$
is kept constant.
\[
\frac{dx^{i}}{d\lambda}=\left\llbracket x^{i},\boldsymbol{H}\right\rrbracket 
=\{x^{i},h_{cl}\}+cp_{0}\{x^{i},\gamma\}\approx\{x^{i},h_{cl}\}
\]
\[
\frac{dp_{i}}{d\lambda}=\left\llbracket p_{i},\boldsymbol{H}\right\rrbracket 
=\{p_{i},h_{cl}\}+cp_{0}\{p_{i},\gamma\}\approx\{p_{i},h_{cl}\}
\]

\subsubsection{Canonical Quantization}

Upon applying the canonical quantization the wave equation based on
the extended Hamiltonian $\boldsymbol{H}$ (\ref{eq:H_for_tau-time}) becomes:
\[
\boldsymbol{H}=\frac{1}{2m}p_{i}p^{i}+mc^{2}+p_{0}c\gamma=0\Rightarrow\left[\frac{1}{2m}\left(\hat{\overrightarrow{P}}
\cdot\hat{\overrightarrow{P}}\right)+mc^{2}\right]\psi=\gamma c\hat{P^{0}}\Psi=i\hbar\gamma\partial_{t}\psi
=i\hbar\partial_{\tau}\psi,
\]
 
\[
i\hbar\partial_{\tau}\psi=\left[-\frac{\hbar^{2}}{2m}\left(\overrightarrow{\nabla}\cdot\overrightarrow{\nabla}\right)+mc^{2}\right]\psi
\]
This is the familiar non-relativistic Schrodinger equation when the
coordinate-time is also the proper-time of the system. In particular,
in the rest frame it implies $\psi\propto exp(-\frac{i(KE+mc^{2})}{\hbar}\tau)$
up to an overall constant factor due to the linear nature of the equation.
Notice that in the above expression the relativistic factor $\gamma$
has been used to switch from the coordinate-time differentiation $\partial_{t}$
to the proper-time differentiation $\partial_{\tau}$. 

In the presence of a non-trivial gravitational field the equation
in the instantaneous rest frame ($\overrightarrow{p}=0$) becomes:
\[
\boldsymbol{H}=mc^{2}-cp^{0}\gamma=0\Rightarrow\boldsymbol{H}=1-\frac{p^{0}}{\phi}=0.
\]
Notice that this equation is the same expression as Equation (\ref{H4tau})
discussed in main text. Here, we continue by considering fluctuations
in the field $\phi$, which is expected to be positive since $m$ and
$\gamma$ are usually positive: 
\begin{equation}
\phi=\left\Vert \phi(t)\right\Vert =\left(\frac{\gamma}{mc}\right)^{-1}
=mc\sqrt{g_{00}}=\sqrt{-g_{\mu\nu}p^{\mu}p^{\nu}},
\label{eq:phi^2}
\end{equation}
upon canonical quantization after symmetrization of the Hamiltonian constraint 
and by dividing with $mc^{2}$ the equation becomes:
\[
\boldsymbol{H}=mc^{2}-\frac{1}{2}\left(p^{0}c\gamma+c\gamma p^{0}\right)
=0\Rightarrow\hat{\boldsymbol{H}}=1-\frac{1}{2}\left(\frac{1}{c\phi}c\hat{P}^{0}+c\hat{P}^{0}\frac{1}{c\phi}\right)=0.
\]
The quantum mechanical wave function $\psi$ is then given by (see the main text related to Equation (\ref{wave function psi(t)})):
\[
\psi(t)=\frac{1}{\mathcal{N}}\sqrt{\phi(t)}\exp\left[-\frac{ic}{\hbar}\intop\phi(t)dt\right].
\]

Now by employing (\ref{eq:phi^2}), the expression becomes:
\[
\psi(t)=\frac{1}{\mathcal{N}}\sqrt{\phi(t)}\exp\left[-\frac{i}{\hbar}mc^{2}\intop_{t_0}^{t}\gamma^{-1}dt\right]
=\frac{1}{\mathcal{N}}\sqrt{\phi(t)}\exp\left[-\frac{i}{\hbar}mc^{2}\left(\tau(t)-\tau_{0}\right)\right].
\]
This shows that this state represents a system with conserved quantity
(see Equation (\ref{eq:Conservation of Energy-Statistically})): 
\[
p^{0}=\left\langle \phi\right\rangle _{\Delta}=\frac{1}{\Delta}\intop_{0}^{\Delta}\phi(t)dt
=mc\frac{1}{\Delta}\intop_{0}^{\Delta}\gamma^{-1}dt=mc.
\]

By applying $c\hat{P}^{0}$ on the state $\psi$, one concludes that
the energy of the particle in the presence of time-fluctuating gravitational
field receives an extra contribution due to the gravitational field:
\vspace{6pt}
\begin{eqnarray*}
c\hat{P^{0}}\psi&=&i\hbar\partial_{\tau}\psi=
i\hbar\gamma\partial_{t}\psi=\left(mc^{2}+i\hbar\gamma\frac{1}{\sqrt{\phi(t)}}\partial_{t}\sqrt{\phi(t)}\right)\psi=\\
&=&\left(mc^{2}-\frac{i\hbar}{2}\partial_{\tau}\ln(\gamma)\right)\psi\approx\left(mc^{2}
+\frac{i\hbar}{2}\frac{1}{\sqrt{g_{00}}}\partial_{t}\ln(g_{00})\right)\psi.
\end{eqnarray*}

For weak gravitational fields when $g_{00}=(1-2U(x)/c^{2})$ this
results in the following expression:
\[
c\hat{P^{0}}\psi=i\hbar\partial_{\tau}\psi=i\hbar\gamma\partial_{t}\psi=\left(mc^{2}-\frac{i\hbar}{c^{2}}\partial_{t}U\right)\psi.
\]
Thus, as long as the fluctuations in the local gravitational potential
$U$ are much smaller than the rest mass of the particle $mc^{4}\gg|\hbar\partial_{t}U|$
then one can expect conservation of the energy that is matching the
rest mass of the particle $E=cp^{0}=mc^{2}$, which is consistent within
the Hamiltonian constraint $\boldsymbol{H}=mc^{2}-cp^{0}\gamma=0$ in the rest frame
($\gamma\approx1$).

Notice that the norm of the state $\psi$ could be set to 1 if the
inner product is defined as: 
\begin{equation}
\left\langle \psi,\psi\right\rangle _{\Delta}=\frac{1}{\Delta}\int dt\psi^{*}\psi
=\frac{1}{\Delta}\int\frac{\phi(t)}{\mathcal{N}^{2}}dt=\frac{1}{\mathcal{N}^{2}\Delta}\int mc\gamma^{-1}dt
=\frac{1}{\Delta}\int d\tau=1, \label{rpspnorm1}
\end{equation}
where $\Delta$ is the measurement window with respect to the rest-frame of the process, 
thus using proper-time, and the normalization factor $\mathcal{N}^{2}$ is set to be equal to 
$p_{0}$, that is\linebreak $\mathcal{N}^{2}=p_{0}=mc$. 

Alternatively, if $\Delta$ is the measurement window with respect
to the laboratory coordinate-time then the norm of the state $\left\Vert \psi\right\Vert ^{2}$
will correspond to an effective constant factor ${\gamma}^{-1}$:
\begin{equation}
\left\Vert \psi\right\Vert ^{2}=\left\langle \psi,\psi\right\rangle _{\Delta}
=\frac{1}{\Delta t}\int \psi^{*}\psi dt=\frac{1}{\mathcal{N}^{2}\Delta t}\int mc\gamma^{-1}dt
=\frac{mc}{\mathcal{N}^{2}}\frac{\Delta\tau}{\Delta t}=\frac{mc}{\mathcal{N}^{2}}{\gamma}^{-1}>0.
\label{rpspnorm_gamma}
\end{equation}
In this respect, the positive norm of the wave-function $(\left\Vert \psi\right\Vert ^{2}>0)$
and the common arrow of time imply $m>0$ and $\gamma>0$ and vice
versa. Thus, in order to have a well defined positive norm it is necessary
to have $\gamma>0$, which then implies positive $p^{0}>0$ since $p^{0}=\gamma\sqrt{-g_{\mu\nu}p^{\mu}p^{\nu}}$.
This implies positivity of the energy $E=cp^{0}>0$ and positivity
of the mass as well since $\sqrt{-g_{\mu\nu}p^{\mu}p^{\nu}}=mc$.
In principle one can consider $m<0$ along with $p^{0}<0$, which will
still guarantee $\gamma>0$. However, coexistence of $m>0$ along
with $m<0$ leads to the proper-time paradox discussed in the section
on the arrow of time. Due to the structure of the state $\psi\propto\gamma^{-1/2}\exp(-\frac{imc^{2}}{\hbar}\tau)$
one can view $\psi^{*}$ as a state corresponding to $m<0$ but it
is better to be viewed as anti-particle with $m>0$ that represents the
time-reversal state of the original process $\psi$ in this case, the
proper-time paradox corresponds to particle-anti-particle annihilation
which results in a photon where the proper-time is ill-defined. Furthermore,
the above expression shows the importance of the positive mass to
guarantee common arrow of time, that is, $\frac{\Delta\tau}{\Delta t}>0$.

\end{paracol}
\reftitle{References}




\end{document}